\documentclass{llncs}
\usepackage[latin1]{inputenc}
\usepackage[english]{babel}
\usepackage{amsmath}
\usepackage{amsfonts}
\usepackage{amssymb}
\usepackage{graphicx}
\usepackage{subfigure}
\usepackage{tikz}
\usepackage{float}
\usetikzlibrary{matrix}

\author{Alexander L\"uck\inst{1} \and Pascal Giehr\inst{2} \and J\"orn Walter\inst{2} \and Verena Wolf\inst{1}}
\institute{Department of Computer Science, Saarland University, Saarbr\"ucken, Germany \and Department of Biological Sciences, Saarland University, Saarbr\"ucken, Germany
}

\title{A Stochastic Model for the Formation of Spatial Methylation Patterns}
\begin{document}
\maketitle

\begin{abstract}
 DNA methylation is an  epigenetic mechanism  whose 
  important role in development has been widely recognized. 
  This epigenetic modification results in heritable changes
   in gene expression not encoded by the DNA sequence.
The underlying mechanisms controlling DNA methylation are only partly understood
and recently different mechanistic models of enzyme activities   responsible for   
 DNA methylation have been proposed.
 Here we extend existing Hidden Markov Models (HMMs) for DNA methylation 
 by describing  the occurrence of spatial  methylation patterns over time
 and propose  several models with different neighborhood dependencies.
 We perform numerical analysis of the HMMs applied to bisulfite sequencing 
 measurements and accurately predict wild-type data. In addition, we 
 find evidence that the enzymes' activities 
 depend  on the left 5' neighborhood but not on the right 3' neighborhood.
\end{abstract}

\section{Introduction}
The DNA code of an organism determines its appearance and behavior by encoding protein sequences.  In addition, there is a multitude of additional mechanisms to control and regulate the ways in which the DNA is packed and processed in the cell  and thus determine the fate of a cell.
One of these mechanisms in  cells is DNA methylation, which is an epigenetic modification  that
occurs at the cytosine (C) bases of eukaryotic DNA. 
 Cytosines  are converted to 5-methylcytosine (5mC) by DNA methyltransferase (Dnmt) enzymes. 
 The neighboring nucleotide of a methylated cytosine is usually guanine (G) and together with the 
 GC-pair on the opposite strand, a common pattern is that 
  two methylated cytosines are located diagonally to each other on opposing DNA strands. 
   DNA methylation at CpG dinucleotides is known to control and mediate gene expression and is therefore essential for cell differentiation and embryonic development.
     In  human somatic cells, approximately 70-80\% of the cytosine nucleotides in CpG dyads are
      methylated on both strands and methylation near gene 
      promoters varies considerably depending on the cell type. 
      Methylation of promoters often correlates with low or no transcription \cite{suzuki2008dna}
      and can be used as a predictor of gene expression \cite{kapourani2016higher}. 
Also significant differences in overall and specific methylation levels exist between different tissue types and between normal cells and cancer cells from the same tissue.
However, the exact mechanism which leads to a methylation of a specific CpG  and the formation of distinct methylation patterns at certain genomic regions is still not fully understood. 
Recently proposed measurement techniques based on hairpin bisulfite sequencing (BS-seq) allow to 
 determine on both DNA strands the level of 5mC at individual CpGs dyads \cite{laird2004hairpin}.
Based on a small hidden Markov model, the probabilities of the  different states of a
 CpG can be accurately estimated (assuming that enough samples per CpG are provided) \cite{aijo2016probabilistic,kyriakopoulos2016h}.

\begin{figure}[t]
	\centering{\includegraphics[scale=0.55]{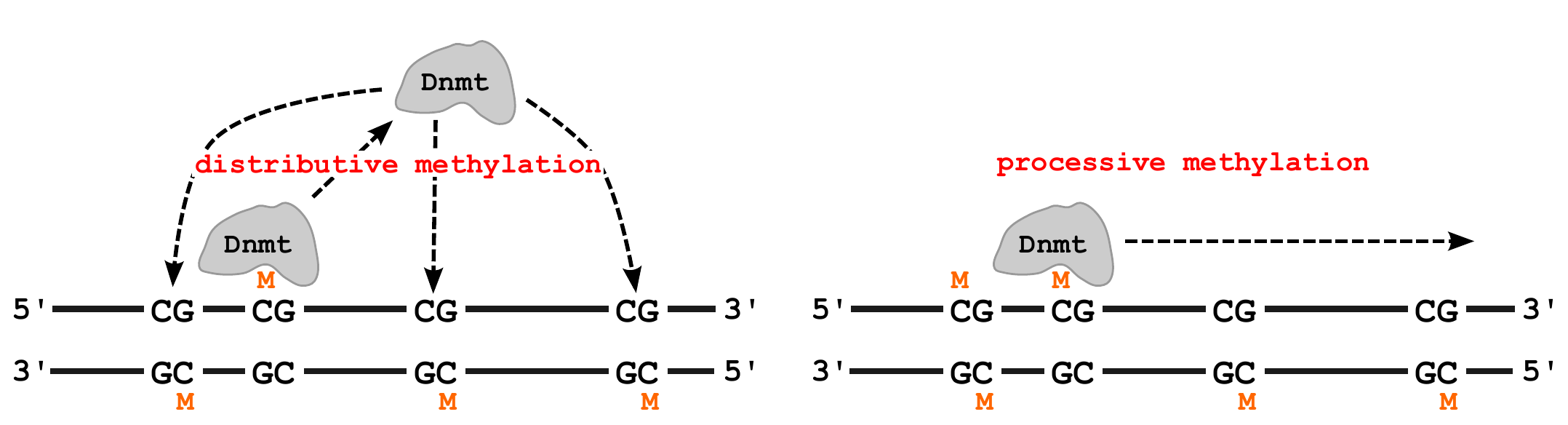}}
	\caption{Dnmts can methylate DNA in a distributive manner, ``jumping'' randomly from one CpG to another or in a processive way where the enzyme starts at one CpG and slides in 5' to 3' direction over the DNA.}
	\label{fig:methylation}
\end{figure}

Mechanistic models for the activity of the different  Dnmts usually distinguish   de novo activities, i.e., adding 
methyl groups at cytosines independent of the methylation state of the opposite strand, 
and maintenance activities, which refers
to the copying of methylation from an existing DNA strand to its newly synthesized partner (containing no
methylation) after replication \cite{hermann2004dnmt1,okano1999dna}. 
Hence, maintenance  methylation is responsible for re-establishment of the same DNA methylation pattern before and after cell replication. 
 A common hypothesis is that the copying of DNA methylation patterns 
 after replication is performed by Dnmt1, 
an enzyme that shows a preference for hemimethylated CpG sites (only one strand is methylated) as they appear after DNA replication. 
Moreover, studies have shown that Dnmt1 is  highly processive and able to methylate long sequences of hemimethylated 
CpGs without dissociation from the target  DNA strand~\cite{hermann2004dnmt1}. 
However, an exact transmission of the methylation information to the next cellular generation is not guaranteed.
The enzymes Dnmt3a and Dnmt3b show equal activities on hemi- and unmethylated DNA and are mainly responsible for de novo methylation, i.e., methylation without any specific preference for the current state of the CpG (hemi- or unmethylated) \cite{okano1999dna}.
 However, by now evidence exists that the activity of the different enzymes is not that exclusive, i.e.,
  Dnmt1 shows to a certain degree also de novo and Dnmt3a/b maintenance methylation activity \cite{arand2012vivo}.
 The way how methyltransferases interact with the DNA and introduce CpG methylation was investigated in many \emph{in vitro} studies.
Basically, one can distinguish between two mechanisms.
A distributive one, where the enzyme periodically binds and dissociates from the DNA, leaping more or less randomly from one CpG to another and a processive one in which the enzyme migrates along the DNA without detachment from the DNA~\cite{gowher2002molecular,holz2010inherent,norvil2016dnmt3b}, as illustrated in
Fig.~\ref{fig:methylation}.
Note that for Dnmt1, for instance, it is reasonable to assume that it is processive in 5' to 3' direction 
since it is linked to the DNA replication machinery.
In particular for the Dnmt3's different hypotheses  about the processivity and neighborhood dependence 
exist \cite{baubec2015genomic,emperle2014cooperative}, but the detailed mechanisms remain elusive.

Several models that  describe the dynamics of the formation of methylation patterns have been proposed.
In the seminal paper of Otto and Walbot, a dynamical model was proposed that assumed independent methylation events for a single CpG. The main idea was to track the frequencies of fully, hemi- and unmethylated CpGs during several cell generations \cite{otto1990dna}. Later, refined models   allowed to distinguish between maintenance and de novo methylation on the parent and daughter strands \cite{genereux2005population,sontag2006dynamics}.
More sophisticated extensions of the original model of Otto and Walbot models have been successfully used to predict \emph{in vivo} data still assuming a neighbor-independent methylation process for a single CpG site \cite{arand2012vivo,giehr2016}.   
However,   measurements indicate that methylation events at a single CpG may depend on the methylation state of neighboring CpGs, which is not captured by these models.

Here, we follow the dynamical HMM approach proposed in \cite{arand2012vivo} where knockout data was used to train a model that accurately predicts wild-type methylation levels for BS-seq data of repetitive elements from mouse embryonic stem cells. 
We extend this model by describing the methylation state of several CpGs instead
of a single CpG and use similar dependency parameters as introduced in   \cite{bonello2013bayesian}.
More specifically, we design different models by combining the activities of the two types of Dnmts and
test for both, maintenance and de novo methylation the  hypotheses illustrated in Fig~\ref{fig:methylation}. 
The models vary according to
the order in which the enzymes act, whether they perform methylation in a processive manner 
or not, and how much their action depends on the left/right CpG neighbor.
We use the same BS-seq data as in \cite{arand2012vivo}, i.e. data where Dnmt1 or Dnmt3a/b was knocked out (KO)
and learn the parameters of the different models. Then, similar as in  \cite{arand2012vivo}, we 
predict the behavior of the measured wild-type (WT), in which both types of enzymes are active, by
designing a combined model that describes the activity of both enzymes and compare the results to
the WT data.

%

We found that all proposed models show a similar behavior in terms of prediction quality such that no model can be declared as the best fit. 
However, our results indicate that Dnmt1   works independently of the methylation state of its neighborhood,
which is in accordance to the current hypothesis that Dnmt1 is linked to the replication machinery
and copies the methylation state on the opposite strand. 
On the other hand, Dnmt3a/b shows a dependency   to the left but no dependency to the right, 
which supports hypotheses of processive or cooperative behavior.

\section{Preliminaries}
\begin{figure}[tb]
\centering{\includegraphics[scale=0.35]{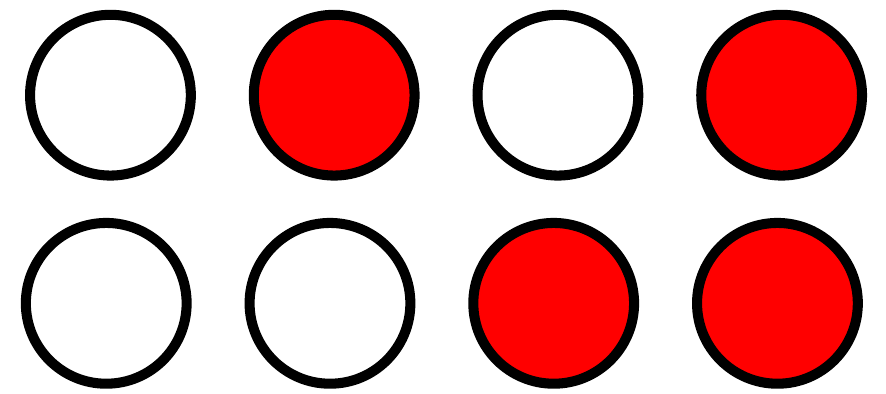}}
\caption{A lattice of length $L=4$ containing all possible states 0, 1, 2 and 3, forming the pattern 0123.}
\label{states}
\end{figure}

Consider a sequence of $L$ neighboring CpG dyads\footnote{The exact nucleotide distance between two
neighboring   dyads is not considered here, but we assume that this distance  is small. For the BS-seq 
data that we  consider, the average distance between two CpGs is 14~bp and the maximal distance is 46~bp.}, 
which is represented as a lattice of length $L$ and width two (for the two strands). 
Each cytosine in the lattice can either be methylated or not, leading to four possible states at each position $l$:
\begin{itemize}
\item \emph{State 0}: Both sites are not methylated.
\item \emph{State 1}: The cytosine on the upper strand is methylated, the lower one not.
\item \emph{State 2}: The cytosine on the lower strand is methylated, the upper one not.
\item \emph{State 3}: Both cytosines are methylated.
\end{itemize}
 A sequence of four CpGs, each of which is in one of the four possible states, is shown in Fig.~\ref{states}.\\
For a system of length $L$ there are in total $4^L$ possibilities to combine the states of individual CpGs. These combinations are called \emph{patterns} in the following. A pattern is denoted by a concatenation of states, e.g. $321$, $0123$ or $33221$.\\
In order to represent the pattern distribution as a vector it is necessary to uniquely assign a reference number to each pattern. A pattern can be perceived as a number in the tetral system, such that converting to the decimal system leads to a unique reference number. After the conversion an additional 1 is added in order to start the referencing at 1 instead of 0.\\
Examples for $L=3$:
\begin{align*}
000~~~&\longrightarrow~~~~~\!1~(=0+1)\\
123~~~&\longrightarrow~~~28~(=27+1)\\
333~~~&\longrightarrow~~~64~(=63+1)
\end{align*}
This reference number then corresponds to the position of the pattern in the respective distribution vector.
\section{Model}
We describe the state of a sequence of $L$ CpGs by a discrete-time Markov chain with pattern distribution 
$\pi(t)$, i.e., the probability of each of the $4^L$ patterns  after $t$ cell divisions. For the initial distribution 
$\pi(0)$, we use the distribution measured in the wild-type when the cells are in equilibrium.
Note that other initial conditions gave very similar results, i.e., the choice of the initial distribution 
does not significantly affect the results. The reason is that also the KO data is   measured after
a relatively high number of cell divisions where the cells are almost in equilibrium.
Transitions between patterns are triggered by different processes: First due to \emph{cell division} the methylation on one strand is kept as it is (e.g. the upper strand), whereas the newly synthesized strand (the new lower strand) does not contain any methyl group. 
Afterwards, methylation is added due to different mechanisms. 
On the newly synthesized strand a site can be methylated if the cytosine 
at the opposite  strand is already methylated (\emph{maintenance}). It is widely accepted that maintenance in form of Dnmt1 is linked to the replication machinery and thus occurs during/directly after the synthesis of the new strand.
Furthermore, CpGs on both strands can be methylated independent of the methylation state of the opposite site (\emph{de novo}). 
The transition matrix $P$ is defined by composition of matrices for   cell division, maintenance and de novo 
methylation of each site.

\subsection{Cell Division}
Depending on which daughter cell is considered after cell replication, the upper ($s=1$) or lower ($s=2$) strand is the parental one after   cell division. Then, the new pattern can be obtained by applying the following state replacements:
\begin{equation}
s=1:~\begin{cases}
0~~~&\longrightarrow~~~0\\
1~~~&\longrightarrow~~~1\\
2~~~&\longrightarrow~~~0\\
3~~~&\longrightarrow~~~1
\end{cases}
\quad\quad s=2:~
\begin{cases}
0~~~&\longrightarrow~~~0\\
1~~~&\longrightarrow~~~0\\
2~~~&\longrightarrow~~~2\\
3~~~&\longrightarrow~~~2
\end{cases}
\label{repl}
\end{equation}
Given some initial pattern with reference number $i$, applying the transformation~\eqref{repl} to each of the 
$L$ positions leads to a new pattern with reference number $j$ (notation: $i\stackrel{\eqref{repl}}{\leadsto} j$). The corresponding transition matrix $D_s \in \{0,1\}^{4^L\times 4^L}$ has the form
\begin{equation}
D_s(i,j)=
\begin{cases}
1,\qquad\text{if~}i\stackrel{\eqref{repl}}{\leadsto} j,\\
0,\qquad\text{else.}
\end{cases}
\end{equation}

\subsection{Maintenance and De Novo Methylation}
\begin{figure}[tb]
\centering{\includegraphics[scale=0.4]{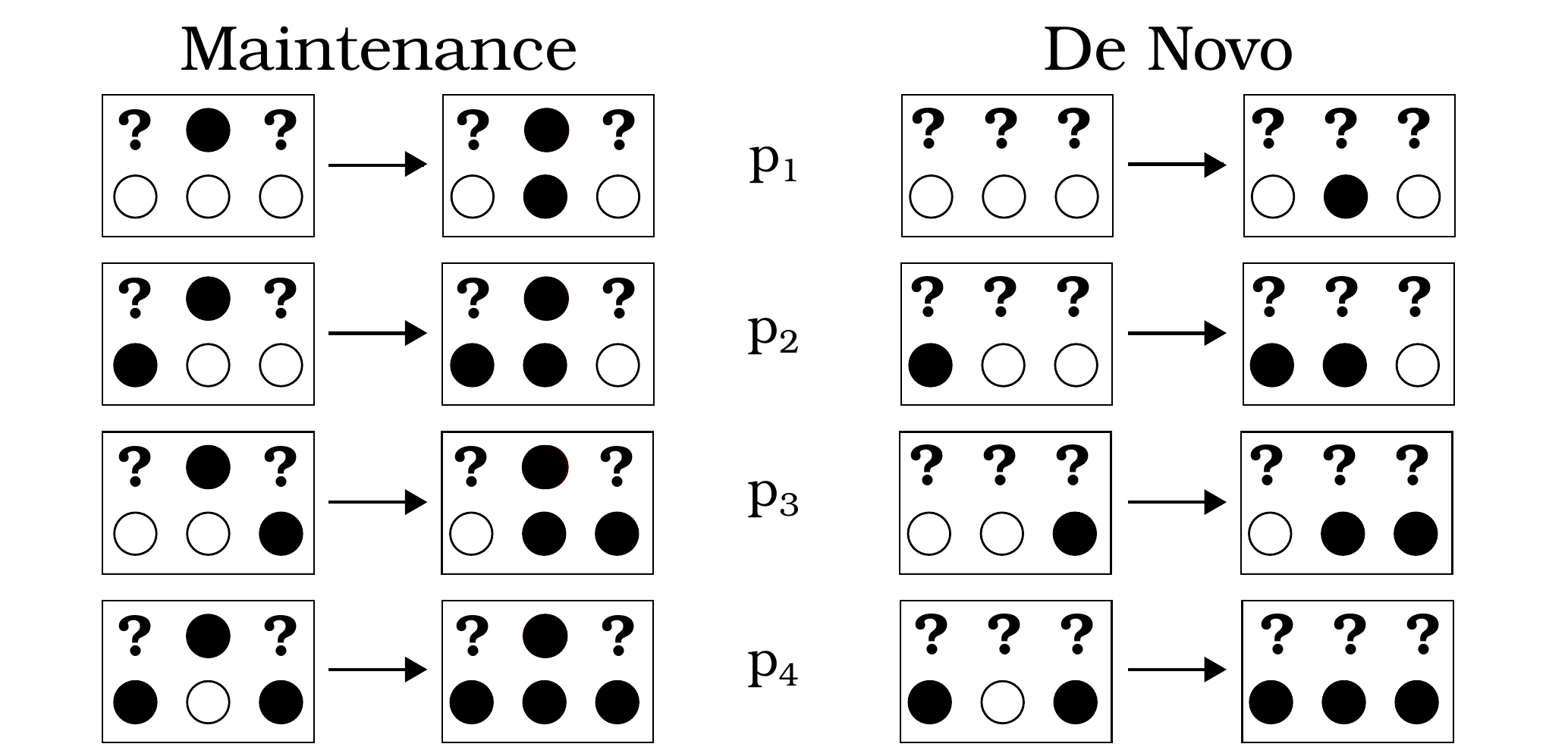}}
\caption{Possible maintenance and de novo transitions depicted for the lower strand, where $\circ$ denotes an unmethylated, $\bullet$ a methylated site and {\bf{?}} a site where the methylation state does not matter. Note that the same transitions can occur on the upper strand.}
\label{fig:Trans}
\end{figure}
For maintenance and de novo methylation, the single site transition matrices are built according to the following rules:\\
Consider at first the (non-boundary) site $l=2,\ldots,L-1$ and its left and right neighbor $l-1$ and $l+1$ respectively. The remaining sites do not change and do not affect the transition. The   probabilities
of the different types of transitions in Fig.~\ref{fig:Trans} have the form
\begin{align}
p_1=&0.5\!\cdot\!(\psi_L+\psi_R)x, \label{pstart}\\
p_2=&0.5\!\cdot\!(\psi_L+\psi_R)x+0.5\!\cdot\!(1-\psi_L),\\
p_3=&0.5\!\cdot\!(\psi_L+\psi_R)x+0.5\!\cdot\!(1-\psi_R),\\
p_4=&1-0.5\!\cdot\!(\psi_L+\psi_R)(1-x),
\end{align}
where $x=\mu$ is the maintenance probability, $x=\tau$ is the de novo probability and $\psi_L,~\psi_R\in[0,1]$ the dependency parameters for the left and right neighbor.\\
A dependency value of $\psi_i=1$ corresponds to a total independence on the neighbor whereas $\psi_i=0$ leads to a total dependence. 
Hence, $\mu$ and $\tau$ can be interpreted as the probability of  maintenance and de novo methylation of a
single cytosine
between two cell divisions assuming independence from   neighboring CpGs.
Moreover, all CpGs that are part of the considered window of the DNA
have the same value for the parameters $\mu$, $\tau$, $\psi_L$, and $\psi_R$,
since in earlier experiments only very small differences have been found between the
methylation efficiencies of nearby CpGs~\cite{arand2012vivo}.

In order to understand the form of the transition probabilities consider at first a case with only one 
neighbor. The probabilities then have the form $\psi x$ if the neighbor is unmethylated and $1-\psi(1-x)$ if the neighbor is methylated. Note that both forms evaluate to $x$ for $\psi=1$, meaning that a site is methylated with probability $x$, independent of its neighbor. For $\psi=0$ the probabilities become $0$ and $1$, meaning that if there is no methylated neighbor the site  cannot be methylated or will be methylated for sure if there is a methylated neighbor respectively.\\
The probabilities for two neighbors are obtained by a linear combination of the one neighbor cases, with $\psi_L$ for the left and $\psi_R$ for the right neighbor,  and an additional weight of $0.5$ to normalize the probability.\\
The same considerations also apply to the boundary sites however there is no way of knowing the methylation states outside the boundaries (denoted by ?). Therefore instead of a concrete methylation state ($\circ$ for unmethylated, $\bullet$ for methylated site) the average methylation density $\rho$ is used to compute the transition probabilities at the boundaries (depicted here for de novo):
\begin{align}
?\circ\circ&\rightarrow\text{?}\bullet\circ\qquad \tilde{p}_1=(1-\rho)\!\cdot\!p_1+\rho\!\cdot\!p_2,\\
?\circ\bullet&\rightarrow\text{?}\bullet\bullet\qquad \tilde{p}_2=(1-\rho)\!\cdot\!p_3+\rho\!\cdot\!p_4,\\
\circ\circ\text{?}&\rightarrow \circ\bullet\text{?}\qquad \tilde{p}_3=(1-\rho)\!\cdot\!p_1+\rho\!\cdot\!p_3,\\
\bullet\circ\text{?}&\rightarrow\bullet\bullet\text{?}\qquad \tilde{p}_4=(1-\rho)\!\cdot\!p_2+\rho\!\cdot\!p_4. \label{pend}
\end{align}
Note that the same considerations hold for maintenance at the boundaries if the opposite site of the boundary site is already methylated.\\
For each position $l$, there are four transition matrices: two for maintenance and two for de novo, namely one for the upper and one for the lower strand in each process.
In order to construct these matrices consider the three positions $l-1$, $l$ and $l+1$, where the transition happens at position $l$. Only the transitions depicted in Fig.~\ref{fig:Trans} can occur. Furthermore the transitions are unique, i.e. for a given reference number $i$ the new reference number $j$ is uniquely determined. For patterns not depicted in Fig.~\ref{fig:Trans} no transition can occur, i.e. the reference number does not change.\\
The   matrix  describing  a maintenance event at position $l$ and strand $s$  has the form
\begin{align}
M_s^{(l)}(i,j)=
\begin{cases}
1,\quad\quad~~~\!&\text{if~}i=j~\text{and}\not\exists j':~ i\leadsto j',\\
1-p,~~~~\!&\text{if~}i=j~\text{and~} \exists j':~ i\leadsto j',\\
p,\quad\quad~~~&\text{if~}i\neq j~\text{and~}i\leadsto j,\\
0,\quad\quad~~~&\text{else,}
\end{cases}
\end{align}
where the probability $p$ is given by one of the Eqs. \eqref{pstart}-\eqref{pend} that describes
the corresponding case and $x=\mu$. 
Note that  $M_s^{(l)}$ depends on $s$ and $l$ since it describes a single transition
from pattern $i$ to pattern $j$, which occurs on a particular strand and at a particular location with probability $p$.
 We define matrices $T_s^{(l)}$ for de novo methylation according to
  the same rules except that $x=\tau$ and the possible transitions are as in Fig.~\ref{fig:Trans}, right. \\
The advantage of defining the matrices position- and process-wise is that different models can be realized by changing the order of  multiplication of these matrices. \\
It is important to note that 5mC can be further modified by oxidation to 5-hydroxymethyl- (5hmC), 5-formyl- (5fC) and 5-carboxyl cytosine(5caC) by Tet enzymes. These modifications are involved in the removal of 5mC from the DNA and can potentially interfere with methylation events. 
However, our data does not capture these modifications and therefore we are not able to consider these modifications in our model.

\subsection{Combination of Transition Matrices}
For all subsequent models it is assumed that first of all 
 cell division happens and 
  maintenance methylation only occurs on the newly synthesized strand given by $s$, 
  whereas de novo methylation happens on both strands.
  Given the mechanisms in Fig.~\ref{fig:methylation}, the two different kinds of methylation events, and the two types of enzymes, there are several possibilities to combine the transition matrices. 
  We consider the following four models, which we found   most reasonable based on the 
  current state of research
  in DNA methylation:
 \begin{enumerate}
\item first processive maintenance and then processive de novo methylation
\begin{equation}
P_s=\prod_{l_1=1}^L M_s^{(l_1)} \prod_{l_2=1}^L T_1^{(l_2)} \prod_{l_3=1}^L T_2^{(l_3)},
\label{Psstart}
\end{equation}
\item first processive maintenance and then de novo in arbitrary order
\begin{equation}
P_s=\frac{1}{(L!)^2}\prod_{l_1=1}^L M_s^{(l_1)} \left(\sum_{\sigma_1\in S_L}\prod_{l_2=1}^L T_1^{(\sigma_1(l_2))}\right)\left( \sum_{\sigma_2\in S_L}\prod_{l_3=1}^L T_2^{(\sigma_2(l_3))}\right),
\end{equation}
\item maintenance and de novo at one position, processive
\begin{equation}
P_s=\prod_{l=1}^L M_s^{(l)} T_1^{(l)} T_2^{(l)},
\end{equation}
\item maintenance and de novo at one position, arbitrary order
\begin{equation}
P_s=\frac{1}{L!}\sum_{\sigma\in S_L}\prod_{l=1}^L M_s^{(\sigma(l))} T_1^{(\sigma(l))} T_2^{(\sigma(l))},
\label{Psend}
\end{equation}
\end{enumerate}
where $S_L$ is the set of all possible permutations for the numbers $1,\ldots,L$.\\
Note that the de novo events on both strands are independent, 
i.e. the de novo events on the upper strand do not influence the de novo events on the lower strand and vice versa, such that $[T_1^{(l)}, T_2^{(l')}]=0$ independent of $\psi_i$\footnote{$[A,B]=AB-BA$ is the commutator of the matrices $A$ and $B$.}. 
Obviously it is important whether maintenance or de novo happens first, since the transition probabilities and the transitions themselves depend on the actual pattern. 
Furthermore in the case $\psi_i<1$ (dependency on right and/or left neighbor) the order of the transitions on a strand matters, i.e. $[M_s^{(l)}, M_s^{(l')}]\neq0$ and $[T_s^{(l)}, T_s^{(l')}]\neq0$ for $l\neq l'$.
The total transition matrix is then given by a combination of the cell division and maintenance/de novo matrices.\\
Recall that we consider two different types of Dnmts, i.e., Dnmt1 and Dnmt3a/b.
If only one type of Dnmt is active (KO data) the matrix has the form
\begin{equation}
P=0{.}5\!\cdot\!(D_1\!\cdot\! P_1 + D_2 \!\cdot\! P_2)
\label{P1}
\end{equation}
and if all Dnmts are active (WT data)
\begin{equation}
P=0{.}5\!\cdot\!(D_1\!\cdot\! P_1\!\cdot\! \tilde{P}_1 + D_2 \!\cdot\! P_2\!\cdot\! \tilde{P}_2),
\label{P2}
\end{equation}
where $P_s$ and $\tilde{P}_s$ have one of the forms \eqref{Psstart}-\eqref{Psend}. This leads to four different models for one active enzyme or 16 models for all active enzymes respectively.
 In the second case $P_s$ represents the transitions caused by Dnmt1 and $\tilde{P}_s$ the 
 transitions caused by Dnmt3a/b. Note that if $\psi_L=\psi_R=1$ all models are the same within each case.
\subsection{Conversion Errors}
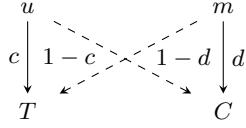
\begin{figure}[bt]
\centering
\begin{tikzpicture}
  \matrix (m) [matrix of math nodes,row sep=3em,column sep=5em,minimum width=3em]
  {
     u & m \\
     T & C \\};
  \path[-stealth]
    (m-1-1) edge node [left] {$c$} (m-2-1)
    (m-1-1) edge [dashed] node [left=0.25cm] {$1-c$} (m-2-2)
    (m-1-2) edge node [right] {$d$} (m-2-2)
    (m-1-2) edge [dashed] node [right=0.25cm] {$1-d$} (m-2-1);
\end{tikzpicture}
\caption{Conversions of the unobservable states $u,m$ to observable states $T,C$ with respective rates.}
\label{ConvErr}
\end{figure}
The actual methylation state of a C cannot be directly observed. 
During BS-seq, with high probability every unmethylated C (denoted by $u$) is converted into Thymine (T) and every
5mC (denoted by $m$) into C.  However, conversion errors may occur and we define their probability  as
$1-c$ and $1-d$, respectively, as shown by the dashed arrows in Fig.~\ref{ConvErr}.
It is reasonable that these conversion errors occur independently  and with approximately identical
probability at each site and thus
 the error matrix for a single CpG takes the form
\begin{equation}
\Delta_1=\begin{pmatrix}
c^2 & c(1-c) & c(1-c) & (1-c)^2 \\ 
c(1-d) & cd & ~~(1-c)(1-d)~~~~ & d(1-c)\\ 
c(1-d) & ~~~~(1-c)(1-d)~~ & cd & d(1-c) \\ 
(1-d)^2 & d(1-d) & d(1-d) & d^2
\end{pmatrix}.
\end{equation}
Due to the independency of the events this matrix can easily be generalized for systems with $L>1$ by recursively using the Kronecker-product 
\begin{equation}
\Delta_L=\Delta_1\otimes\Delta_{L-1}\qquad\text{for~}L\geq 2.
\end{equation}
Hence, $\Delta_L$ gives the probability of observing a certain sequence of C and T nucleotides
for each given   unobservable methylation pattern. 
In order to compute the likelihood $\hat{\pi}$ of the observed BS-seq data, we therefore first compute
 the transient distribution $\pi(t)$ of the underlying Markov chain at the corresponding time
instant\footnote{The number of cell divisions is estimated from the time
 of the measurement since these cells
 divide once every 24 hours.}  $t$ by solving 
 \begin{equation}\label{eq:stst}
{\pi}(t)=\pi(0)\cdot P^t
\end{equation}
and then multiply the distribution  of the unobservable patterns with the error matrix.
\begin{equation}\label{eq:distr}
\hat{\pi}=\pi(t)\cdot \Delta_L.
\end{equation}
Note that  this yields a hidden Markov model with emission probabilities $\Delta_L$.
In the following the values for $c$ were chosen according to \cite{arand2012vivo}. Since the value
for $d$ was not determined in \cite{arand2012vivo},  we 
  measured the conversion rate $d=0.94$ in an independent experiment under comparable conditions (data not shown).
 
\subsection{Maximum Likelihood Estimator}
In order to estimate the parameters $\theta=(\mu,\psi_L,\psi_R,\tau)$, we employ a Maximum (Log)Likelihood Estimator~(MLE)
\begin{equation}
\hat\theta= \arg\max_\theta \ell(\theta),\quad  \ell(\theta)=\sum_{j=1}^{4^L}\log(\hat{\pi}_j(\theta))\!\cdot\! N_j,
\end{equation}
where $\hat{\pi}$ is the pattern distribution obtained from the numerical solution of \eqref{eq:stst} and
\eqref{eq:distr} for a given time $t$ and $N_j$ is the number of occurrences of pattern $j$ in the measured data. The parameters $\theta=\hat{\theta}$ are chosen in such a way that $\ell$ is maximized.
Visual inspection of all two dimensional cuts of the likelihood landscapes showed only a single
local maximum.\\
We employ the MLE twice in order to estimate the parameter vector $\hat{\theta}_1$ for Dnmt1 from the 3a/b DKO (double knockout) data and the vector $\hat{\theta}_{3a/b}$ for Dnmt3a/b from the Dnmt1 KO data, where   transition matrix~\eqref{P1} is used. The corresponding time instants are $t=26$ for the 3a/b DKO data and $t=41$ for the 1KO data.\\
We approximate the standard deviations of the estimated parameters $\hat{\theta}$ as follows: Let $\mathcal{I}(\hat{\theta})=\mathbb{E}[-\mathcal{H}(\hat{\theta})]$ be the expected Fisher information, with the Hessian $\mathcal{H}(\hat{\theta})=\nabla\nabla^\intercal \ell(\hat{\theta})$. The inverse of the expected Fisher information is a lower bound for the covariance matrix of the MLE such that we can use the approximation $\sigma(\hat{\theta})\approx\sqrt{\text{diag}(-\mathcal{H}(\hat{\theta}))}$.\\
A prediction for the wild-type can be computed by combining the estimated vectors such that in
the model both types of enzymes are active. 
For this, we insert $\hat{\theta}_1$ in $P_s$ and $\hat{\theta}_{3a/b}$ in $\tilde{P}_s$ in~\eqref{P2} to obtain
 the transition matrix for the wild-type.

\section{Results} 
For our analysis we 
 focused at the single copy genes Afp (5 CpGs) and Tex13 (10 CpGs) as well as the repetitive elements IAP (intracisternal A particle) (6 CpGs), L1 (Long interspersed nuclear elements) (7 CpGs) and mSat (major satellite) (3 CpGs). Repetitive elements occur in multiple copies and are dispersed over the entire genome. Therefore they allow capturing an averaged, more general behavior of methylation dynamics. 
 If a locus contains more than three CpGs, the analysis is done for all sets of three adjacent sites independently, in order to keep  computation times short and memory requirements low.
In the sequel, we mainly focus on the estimated dependency parameters $\psi_L$ and $\psi_R$
and on the prediction quality of the different models.

\begin{figure}[bt]
\centering
\subfigure{\includegraphics[scale=0.45]{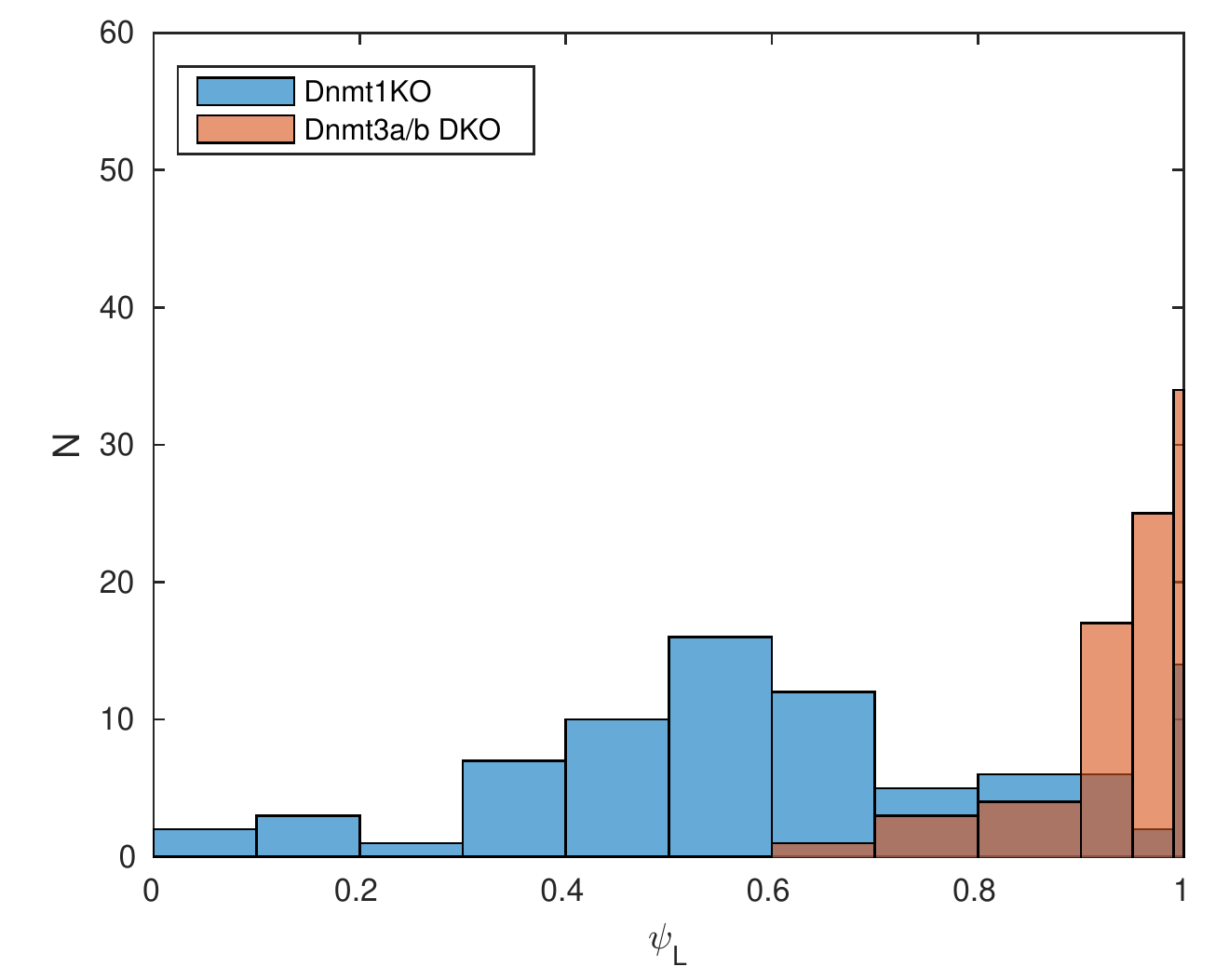}}\hspace{0.1cm}
\subfigure{\includegraphics[scale=0.45]{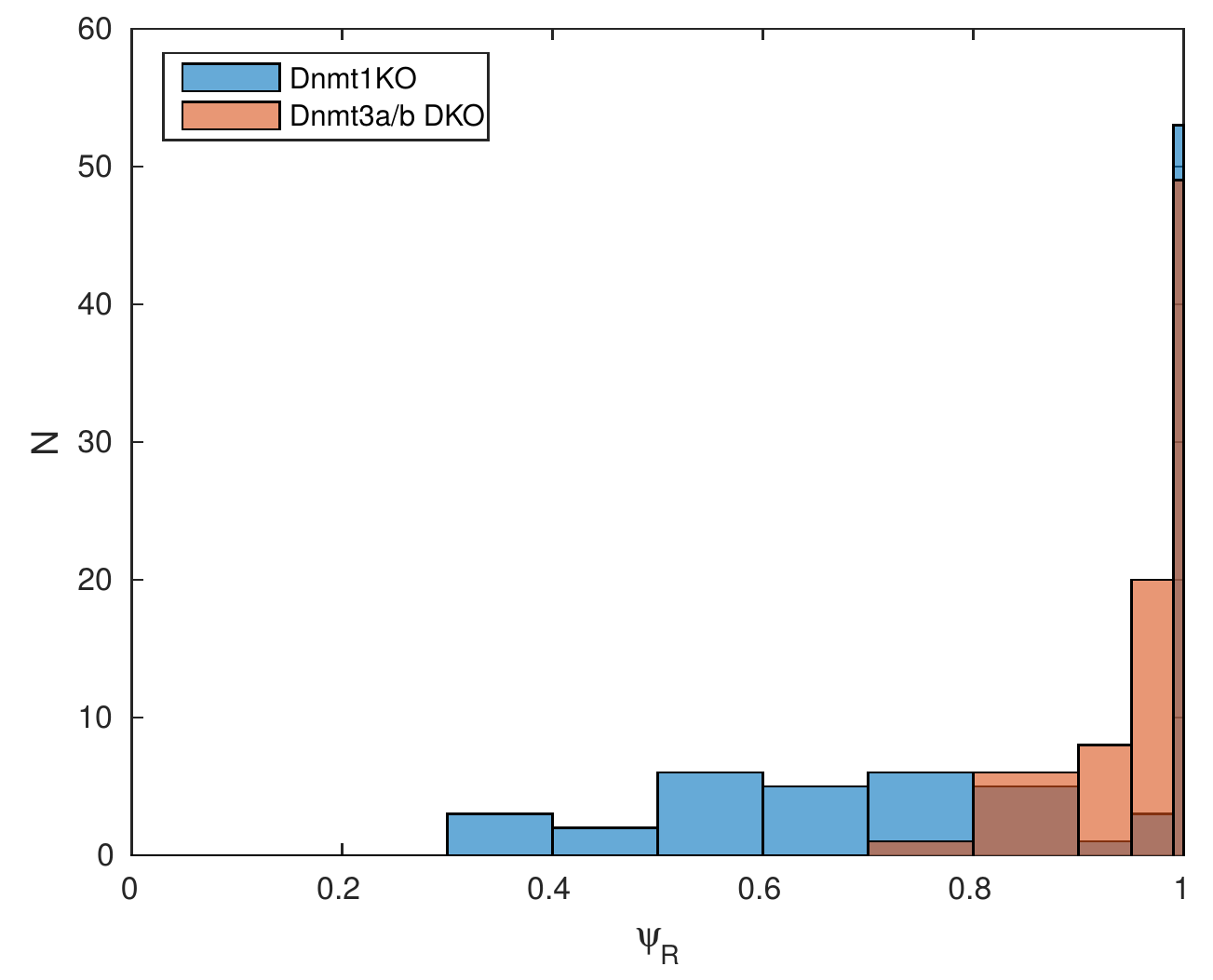}}
\caption{Histograms for the estimated dependency parameters $\psi_L$ and $\psi_R$ for all sets of three adjacent CpGs in all loci and for all suggested models.}
\label{Histo}
\end{figure}

The estimates for all the available KO data and all suggested models obtained   using the transition matrix in Eq.~\eqref{P1} are summarized as histograms in Fig.~\ref{Histo}. 
Because of the different possibilities to combine the four different models in Eq.~\eqref{Psstart}-\eqref{Psend}
and because of the different loci considered,
in total there are 84 estimates for each KO data set.
We plot the number of occurrences $N$ of $\psi_L$ (left) and  $\psi_R$ (right) in different ranges for both sorts of KO data (Dnmt1KO and Dnmt3a/b DKO). 

\begin{figure}[H]
\centering
\subfigure[Afp]{\includegraphics[scale=0.39]{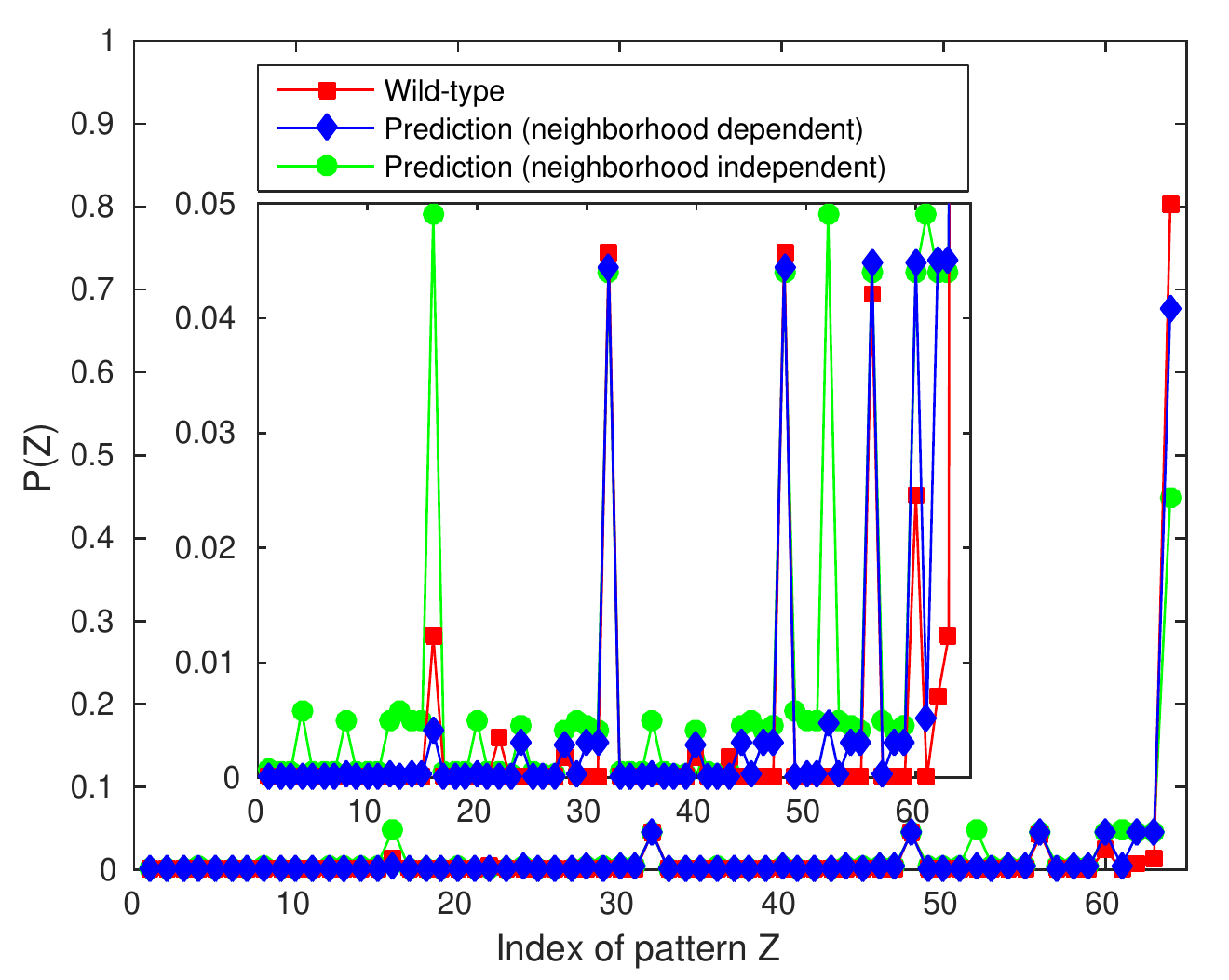}}\hspace{0.5cm}
\subfigure[L1]{\includegraphics[scale=0.39]{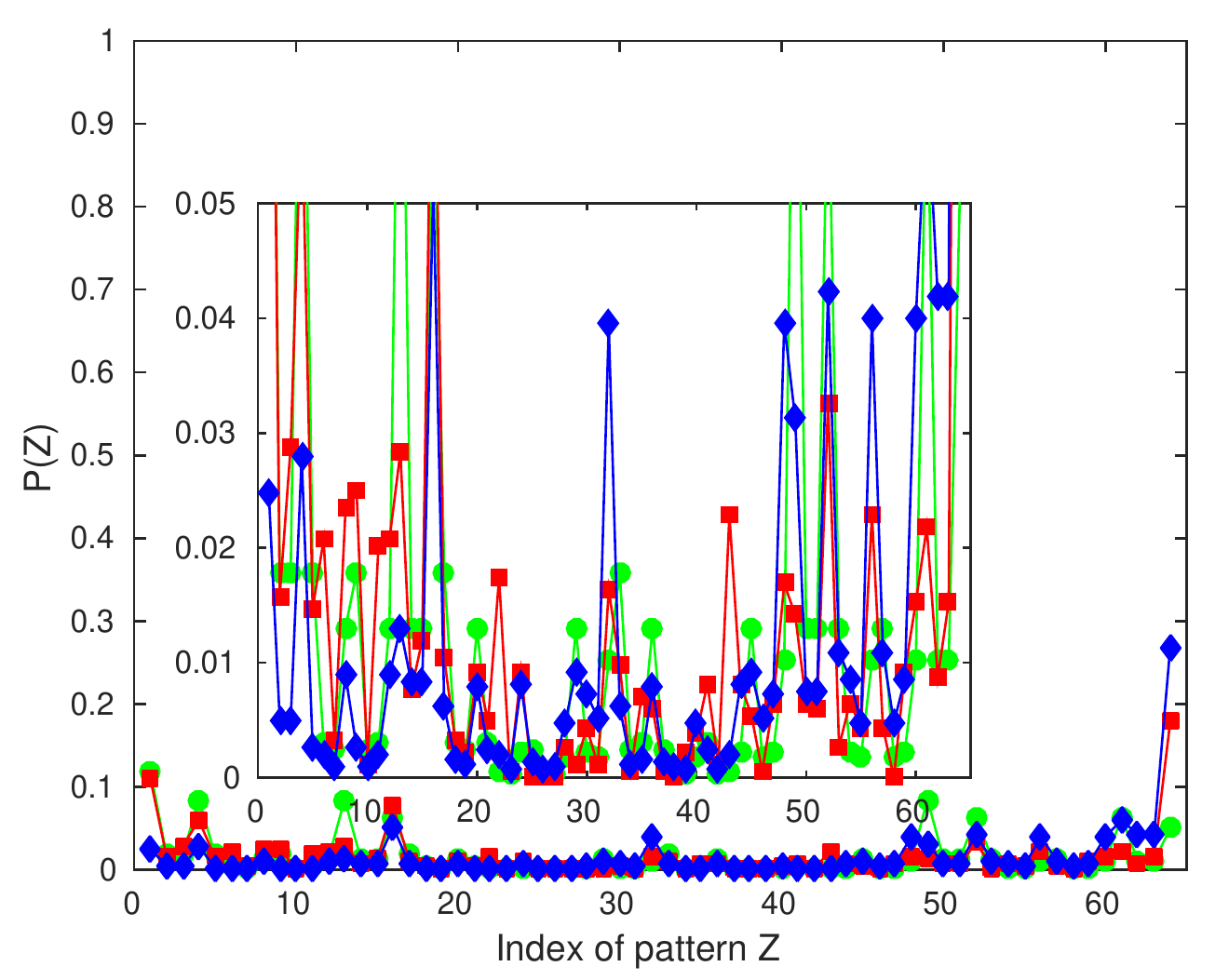}}\\[-0.22cm]
\subfigure[IAP]{\includegraphics[scale=0.39]{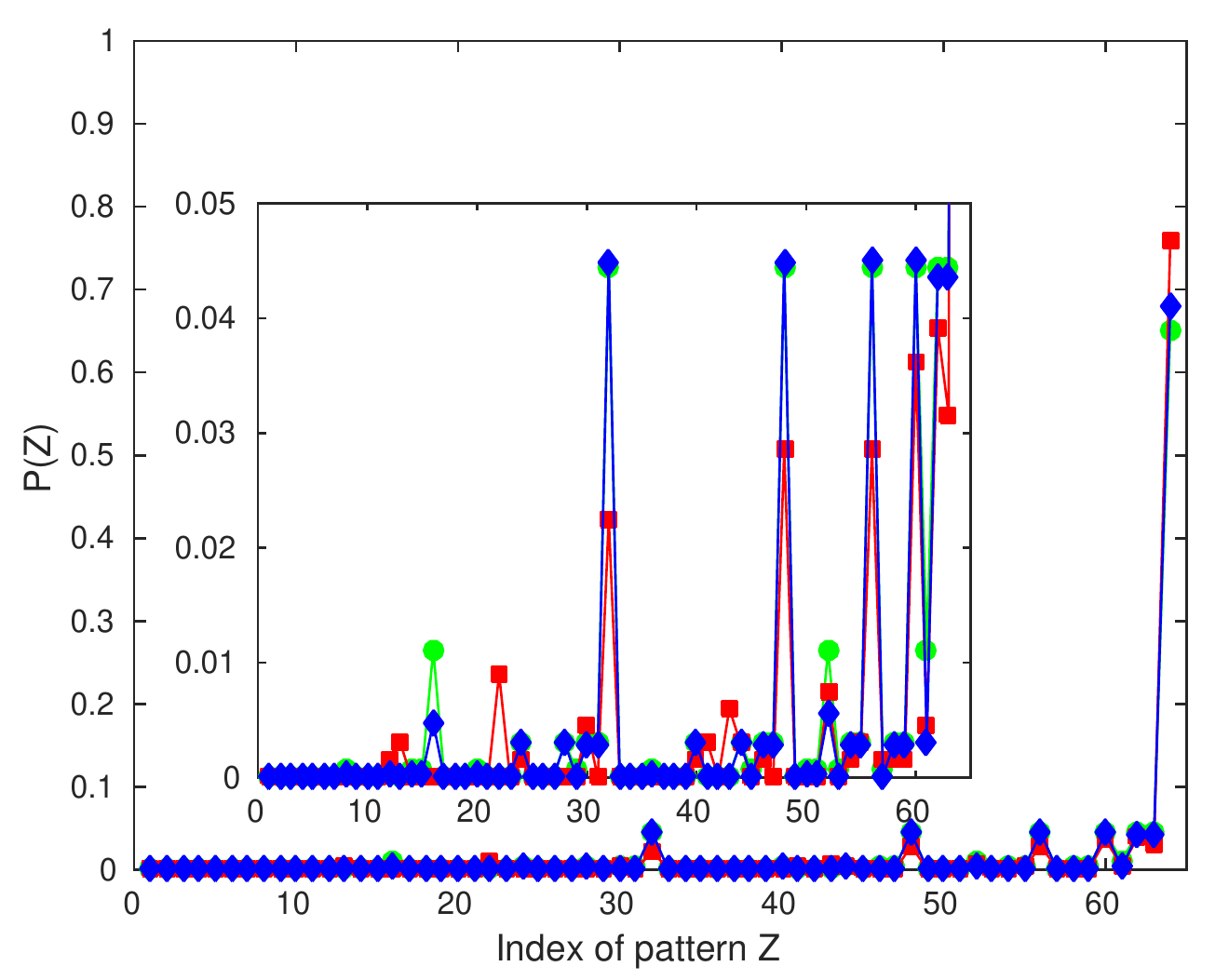}}\hspace{0.5cm}
\subfigure[Tex13]{\includegraphics[scale=0.39]{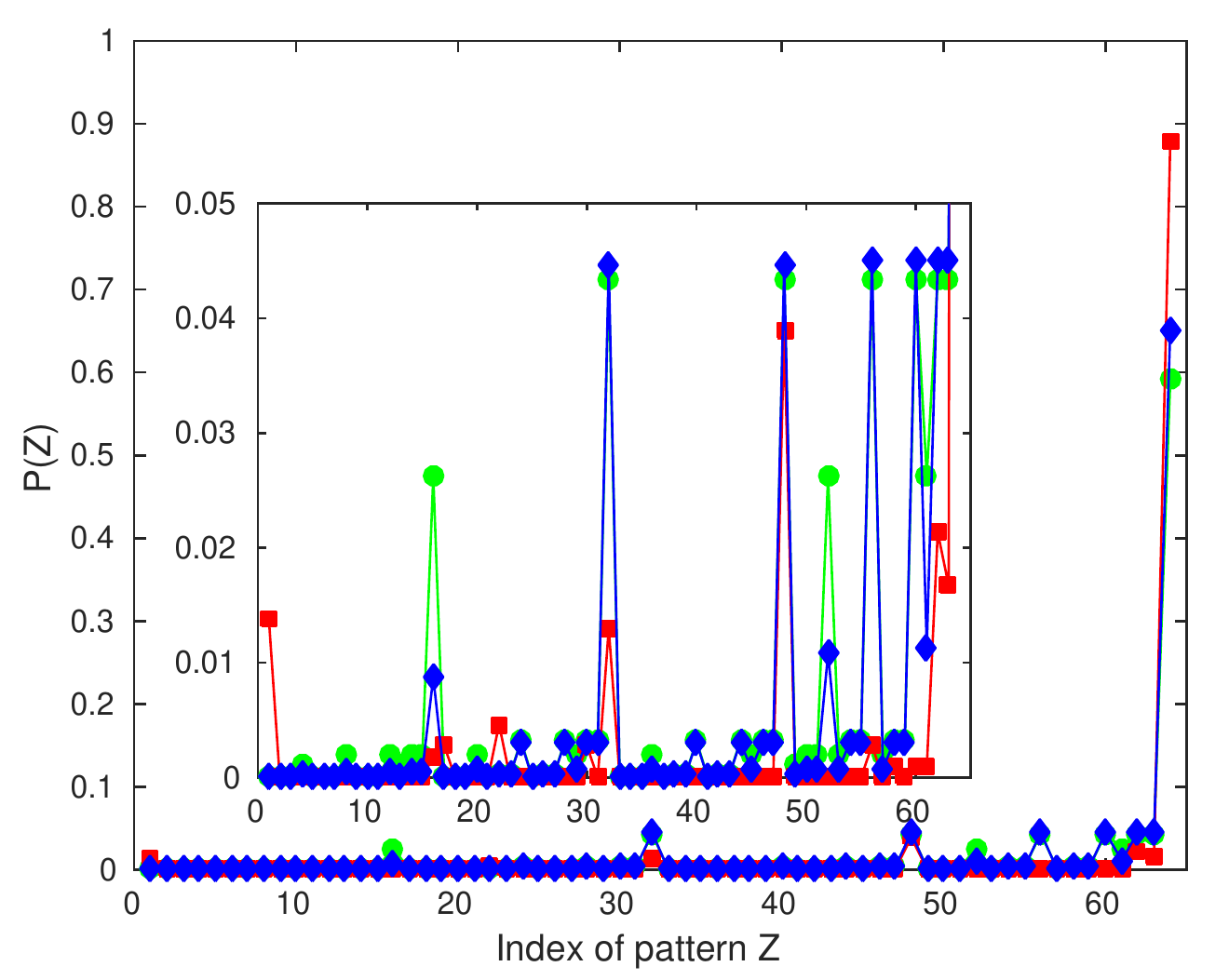}}\\[-0.22cm]
\subfigure[mSat]{\includegraphics[scale=0.39]{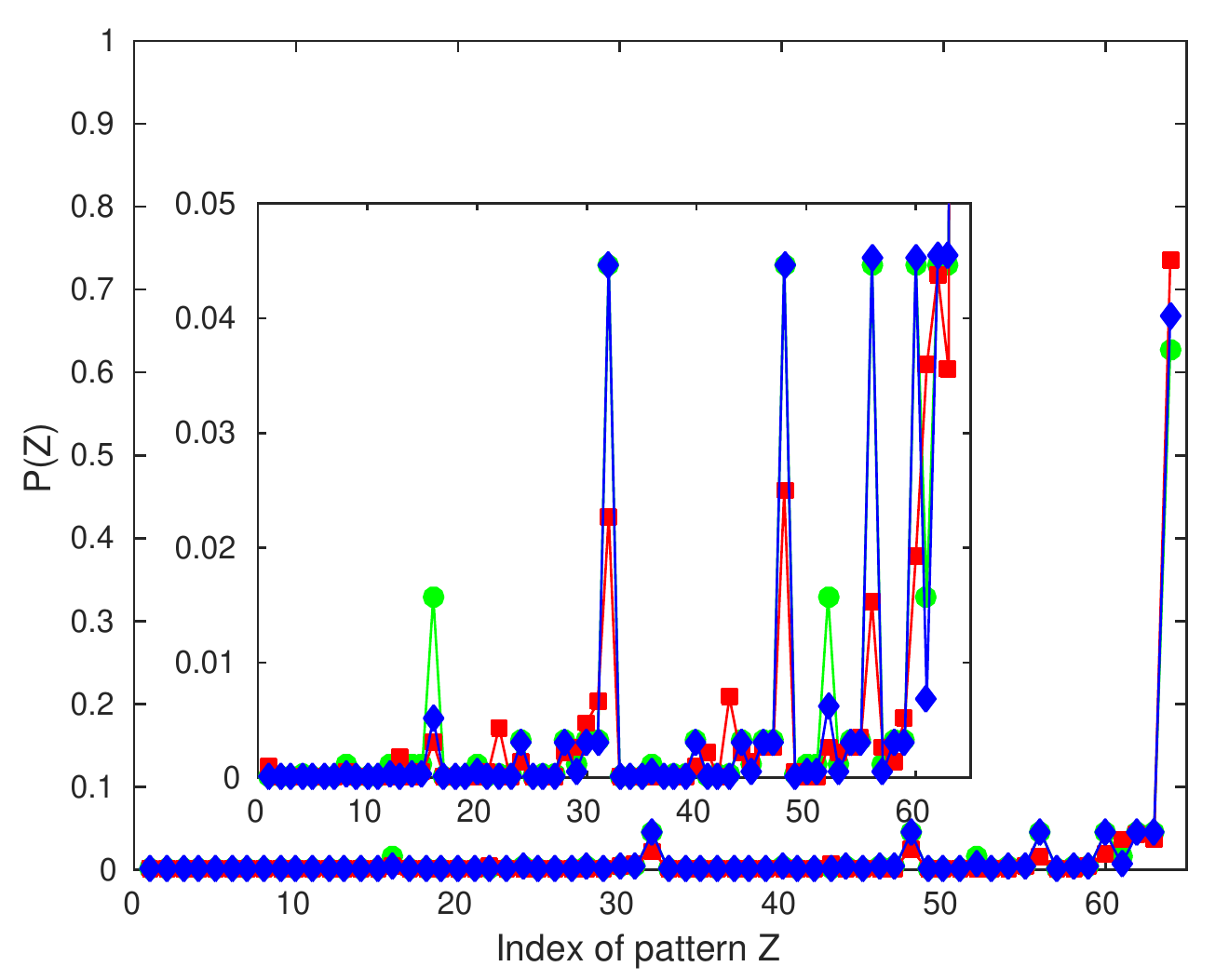}}\\[-0.35cm]
\caption{The figures show an example for the predicted (neighborhood dependent and neighborhood independent) and the measured pattern distribution for each locus. The inset shows a zoomed in version of the distribution.\label{Fig:App:Prediction_loci}}
\end{figure}

The  estimates of $\psi_L$  spread over the whole interval $[0,1]$ while 
in the  case of $\psi_R$, nearly all estimates are  larger than $0.99$ and only in a few cases the dependency parameter is significantly smaller than $1$. Hence, in most cases the methylation probabilities are independent of the right neighbor for both Dnmt1KO and Dnmt3a/b DKO.
For $\psi_L$
the dependency parameter in the Dnmt3a/b DKO case occurs more often close to 1, meaning that the transitions induced by Dnmt1 have little to no dependency on the left neighbor.
On the other hand for Dnmt1KO the dependency parameter occurs more often at smaller values giving evidence that there is a dependency on the left neighbor for the activity of Dnmt3a/b.
Note that all models show a similar behavior in terms of the dependency parameters for a given locus or position within a locus respectively, i.e. either $\psi_i\approx1$ or $\psi_i<1$ for all models.
The difference between the behaviors at different loci and positions may be explained by explicitly including the distances between the CpGs and is planned as  future work.
\\
Since $\psi_R$ is usually close to 1 a smaller model with only three parameters $\theta=(\mu,\psi,\tau)$ can be proposed, where $\psi$ is a dependency parameter for the left neighbor. This model can either be obtained by fixing $\psi_R=1$ in the original model and setting $\psi=\psi_L$ or by redefining the transition probabilities to $\psi x$ if the left neighbor is unmethylated and $1-\psi(1-x)$ if the left neighbor is methylated. 
In that case $\psi$ and $\psi_L$ are related via $\psi=0.5(\psi_L +1)$. Note that both versions yield the same results.\\
In order to check whether there is a significant difference in the original and the smaller model, we performed a Likelihood-ratio test with the null hypothesis that the smaller model is a special case of the original model. Since the original model with more parameters is always as least as good as the smaller model, our goal is to check in which cases the smaller model is sufficient. Indeed if $\psi_R$ was estimated to be approximately 1 the Likelihood-ratio test indicates that the smaller model is sufficient (p-value $\approx 1$). On the other hand, for the few cases where
 $\psi_R$ differs significantly from 1 the original model has to be used (p-value $< 0.01$).\\
As a next step we used the estimated parameters from the KO data to predict the WT data.
The models from Eq.\eqref{Psstart}-\eqref{Psend} are referred to as \emph{Models 1-4}. For the prediction, the notation $(x,y)$ is used to refer to Model $x$ for the Dnmt3a/b DKO (only Dnmt1 active) and Model $y$ for the Dnmt1KO case (only Dnmt3a/b active).
One instance of the prediction, for which Model 1 was used for both Dnmt1KO and Dnmt3a/b DKO, i.e. $(1,1)$, are shown in Fig.~\ref{Fig:App:Prediction_loci}. Note that all wild-type predictions yielded a very similar accuracy.
We list the corresponding estimations for the parameters for an example of a single copy gene  (Afp) and a repetitive element (L1) in Tab.~\ref{Tab:Paras}.
While the standard deviation of the estimated parameters for $\mu$ is always of the order $10^{-2}$ and for $\tau$ of order $10^{-3}$, it is usually of order $10^{-2}$ for $\psi_i$. Depending on the model, locus and position, standard deviations up to order $10^{-1}$ may occur for the dependency parameters in a few cases.

In Fig.~\ref{Fig:App:Prediction_loci} the predictions for the pattern distribution together with the WT pattern distribution and a prediction from the neighborhood independent model ($\psi_L=\psi_R=1$) for all loci are shown in the main plot. 
As an inset the distributions are shown on a smaller scale to display   small deviations. With the exception of patterns 0 and 64 (which corresponds to no methylation/full methylation of all sites) in L1 and pattern 64 in all loci, where the difference between WT and the numerical solution is about $10\%$, the difference is always small ($<5\%$) as seen in the insets.

In general all 16 models show a similar performance for all loci and positions in terms of accuracy of the prediction.
On the large scale the differences are not visible and even for the smaller scale the differences are small, as shown for mSat in Fig.~\ref{Fig:App:Prediction_mod1}.
This is in accordance to the corresponding Kullback-Leibler divergences
\begin{equation}
KL=\sum_{j=1}^{4^L} \pi_j(\text{WT})\log\left(\frac{\pi_j(\text{WT})}{\pi_j(\text{pred})}\right)
\end{equation}
that we list in Tab.~\ref{Tab:App:KL}.
The difference in $KL$ between the ``best" and the ``worst" case is about $0.01$.
The mean and standard deviation for $KL$ was obtained via bootstrapping of the wild-type data ($10.000$ bootstrap samples for each model).
Since no confidence intervals of the parameters are included, this standard deviation can be regarded as a lower bound.
However, even with these lower bounds the intervals of $KL$ overlap for all models, such that no model can be favorized.


\renewcommand{\arraystretch}{1.25}
\begin{table}[H]
\caption{Estimated parameters for the KO data and model based on Eq.~\eqref{Psstart} for the loci Afp and L1 with sample size $n$.}
\label{Tab:Paras}
\centering
\begin{tabular}{|c|c|c|c|c|c|c|}
\hline 
KO & $\mu$ & $\psi_L$ & $\psi_R$ & $\tau$ & $n$ & Locus \\ 
\hline  
Dnmt1 & $0.452\pm0.062$ & $0.383\pm0.076$ & $1.000\pm0.094$ & $0.091\pm0.016$ & 134& Afp \\ 
Dnmt3a/b & $0.990\pm0.003$ & $0.984\pm0.011$ & $1.000\pm0.006$ & $10^{-10}\pm0.011$ & 186& Afp \\ 
Dnmt1 & $~0.334\pm0.051~$ & $~0.576\pm0.067~$ & $~1.000\pm0.122~$ & $~0.038\pm0.004~$ & 1047& L1 \\ 
Dnmt3a/b & $0.789\pm0.037$ & $1.000\pm0.038$ & $0.984\pm0.045$ & $10^{-10}\pm0.002$ & 805& L1 \\
\hline 
\end{tabular} 
\end{table}
\renewcommand{\arraystretch}{1}
\vspace*{-1.25cm}

\pagebreak
\begin{figure}[H]
\centering
\subfigure[(1,1)]{\includegraphics[scale=0.4]{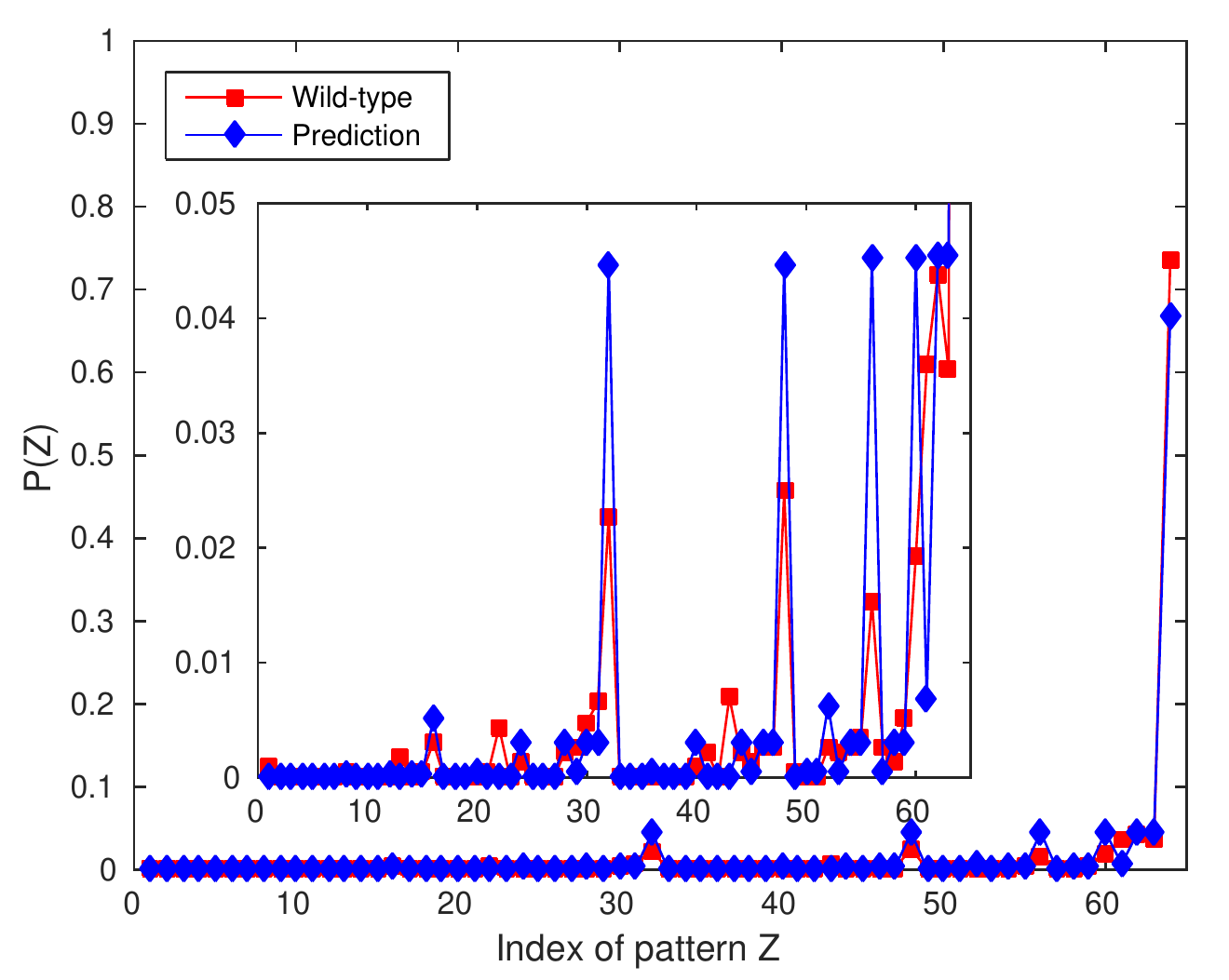}}\hspace{0.5cm}
\subfigure[(1,2)]{\includegraphics[scale=0.4]{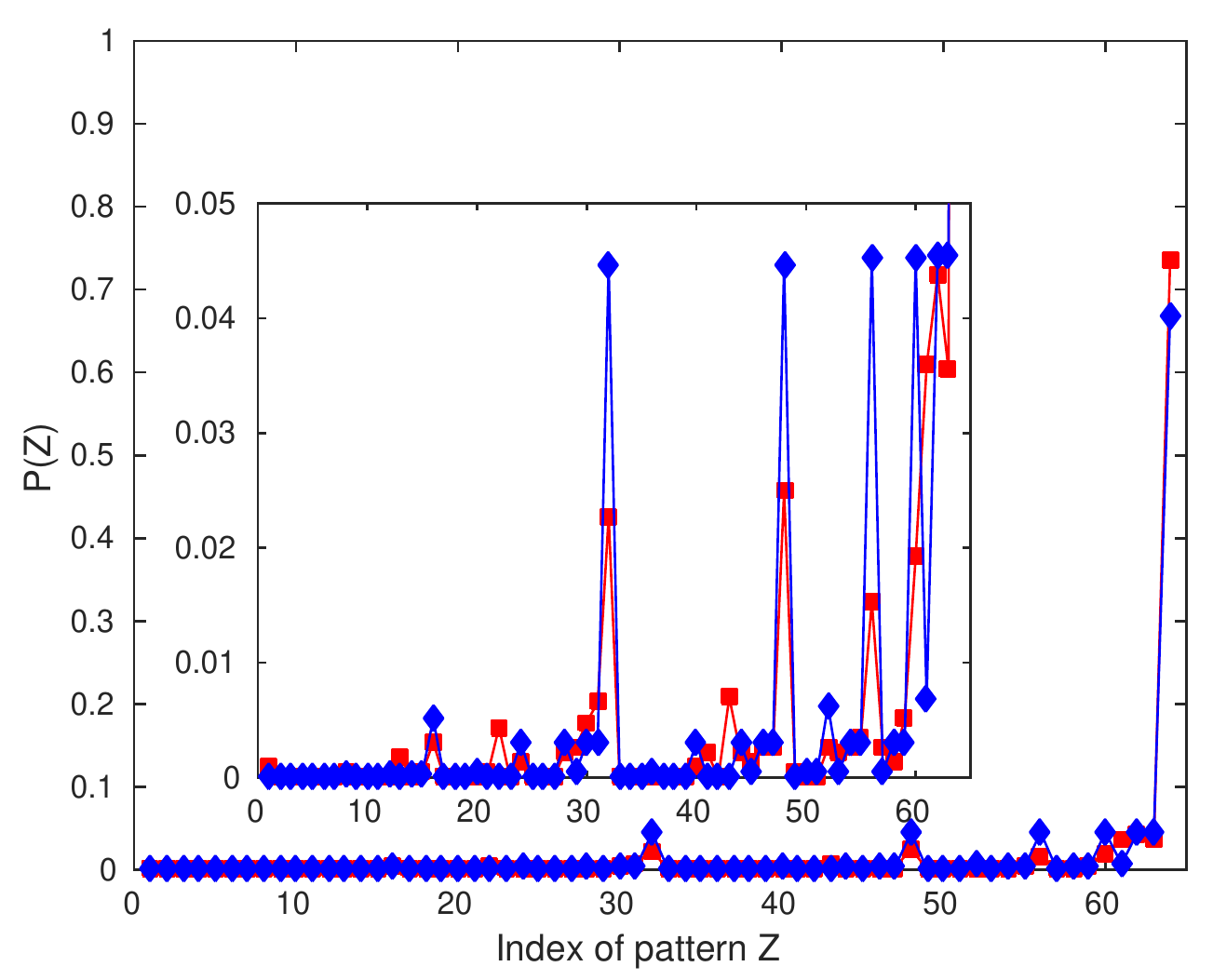}}\\[-0.2cm]
\subfigure[(1,3)]{\includegraphics[scale=0.4]{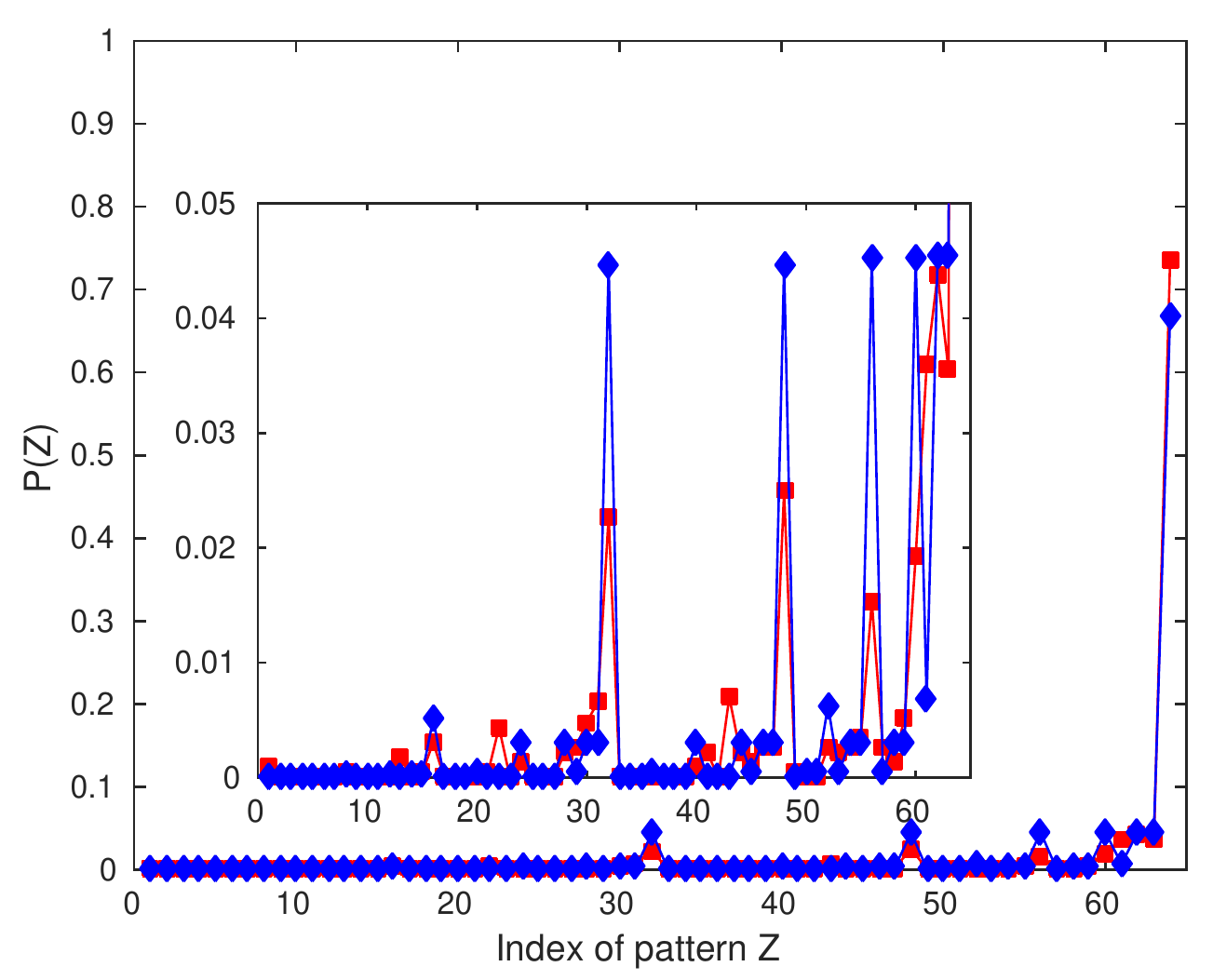}}\hspace{0.5cm}
\subfigure[(1,4)]{\includegraphics[scale=0.4]{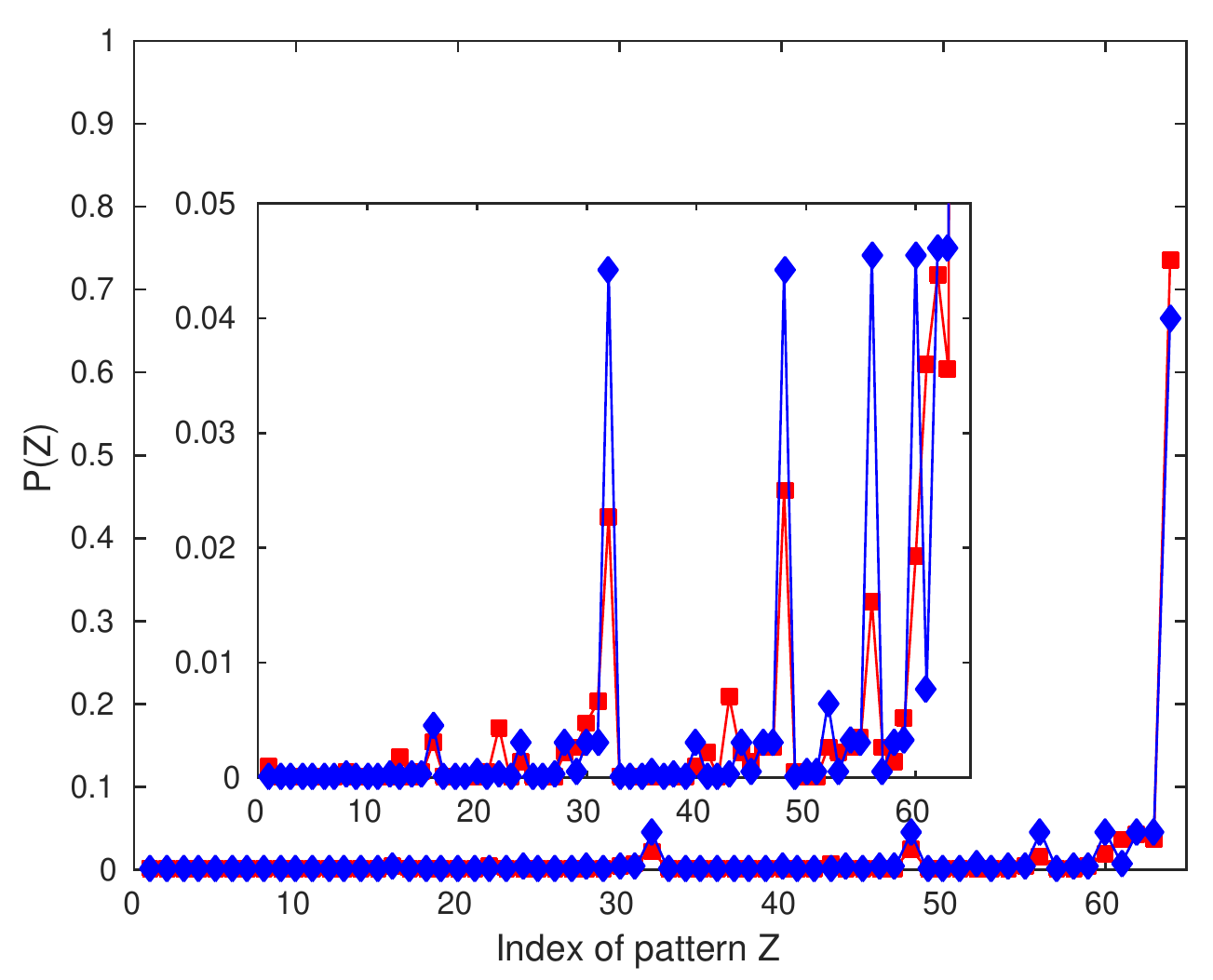}}\\[-0.2cm]
\subfigure[(2,1)]{\includegraphics[scale=0.4]{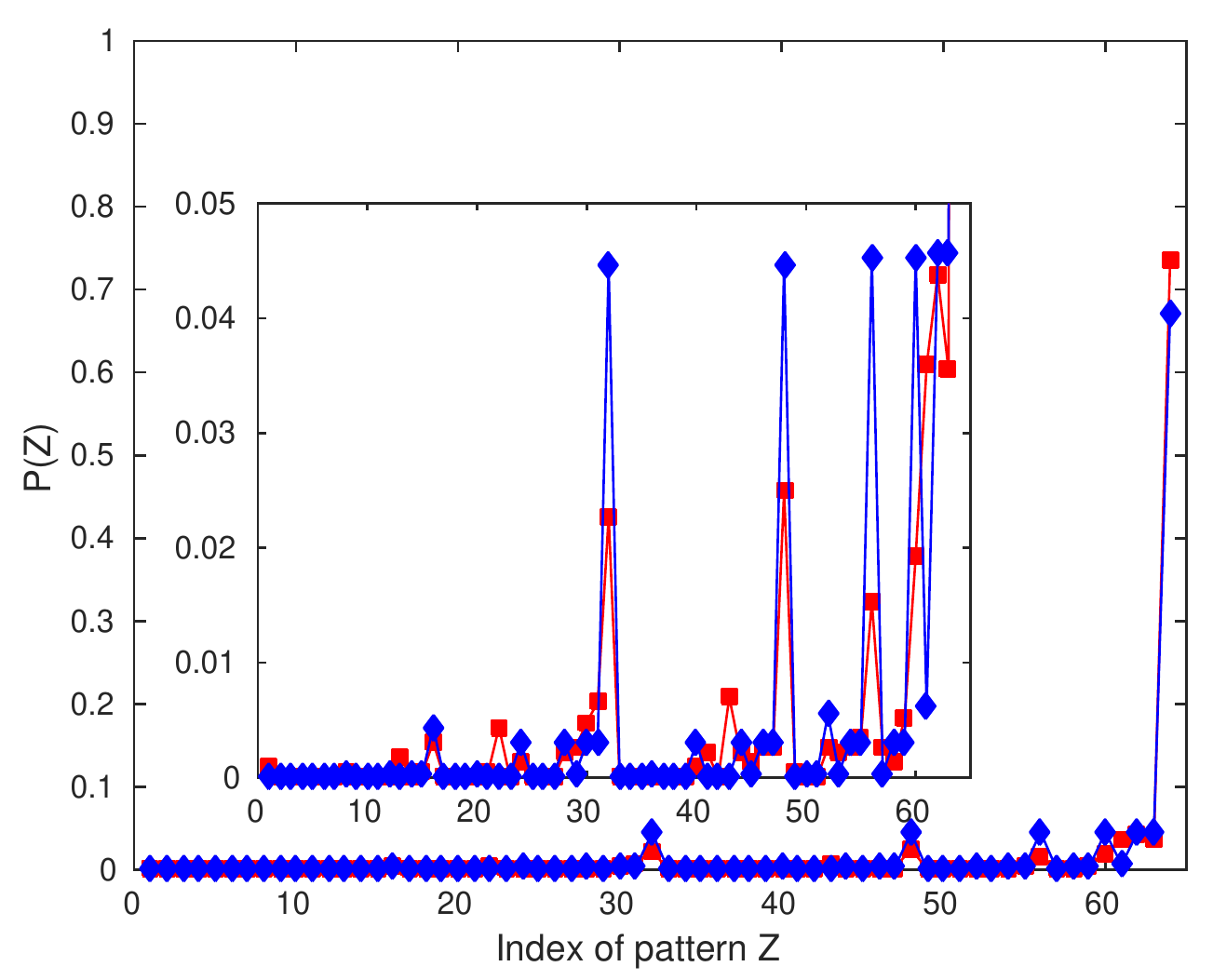}}\hspace{0.5cm}
\subfigure[(2,2)]{\includegraphics[scale=0.4]{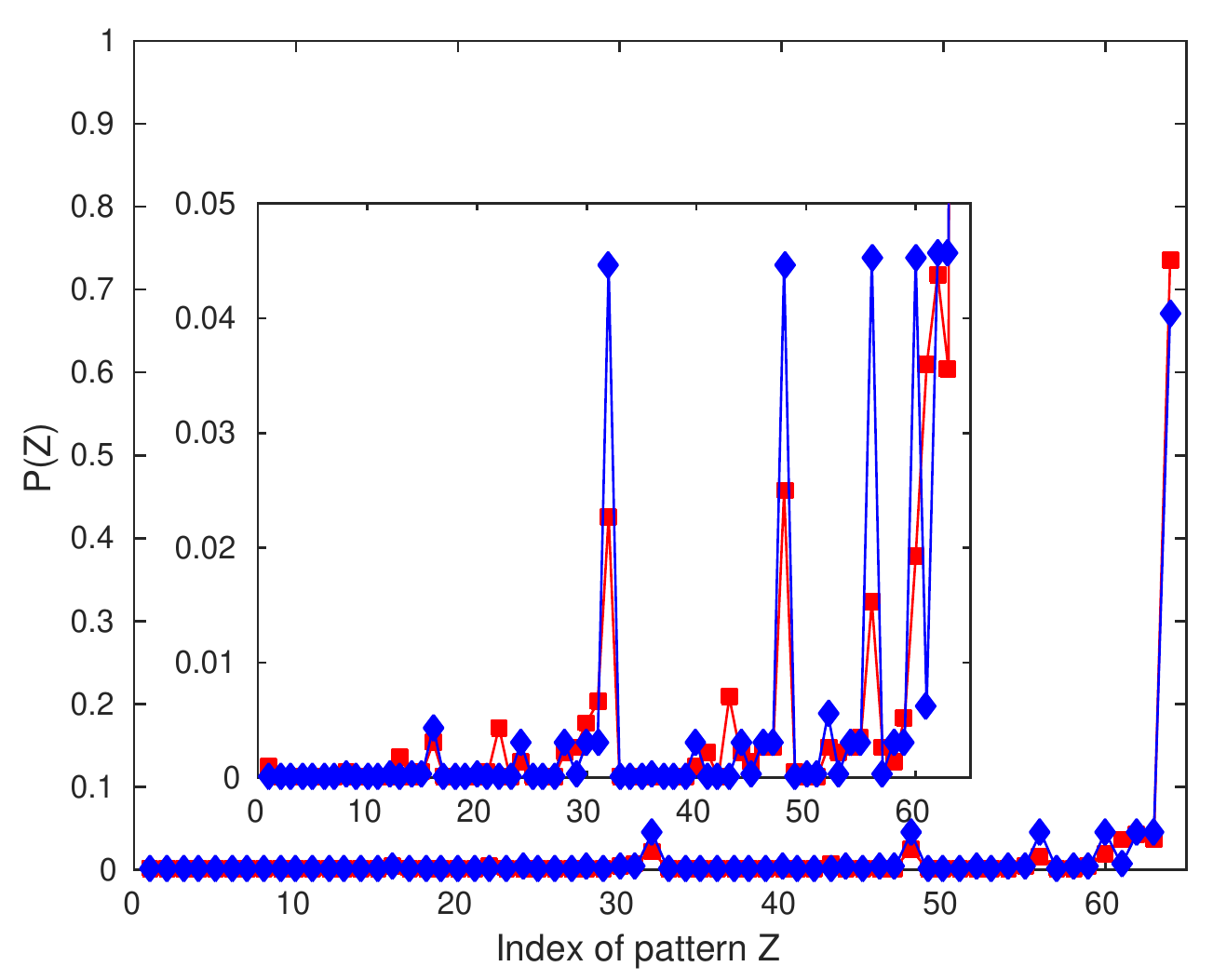}}\\[-0.2cm]
\subfigure[(2,3)]{\includegraphics[scale=0.4]{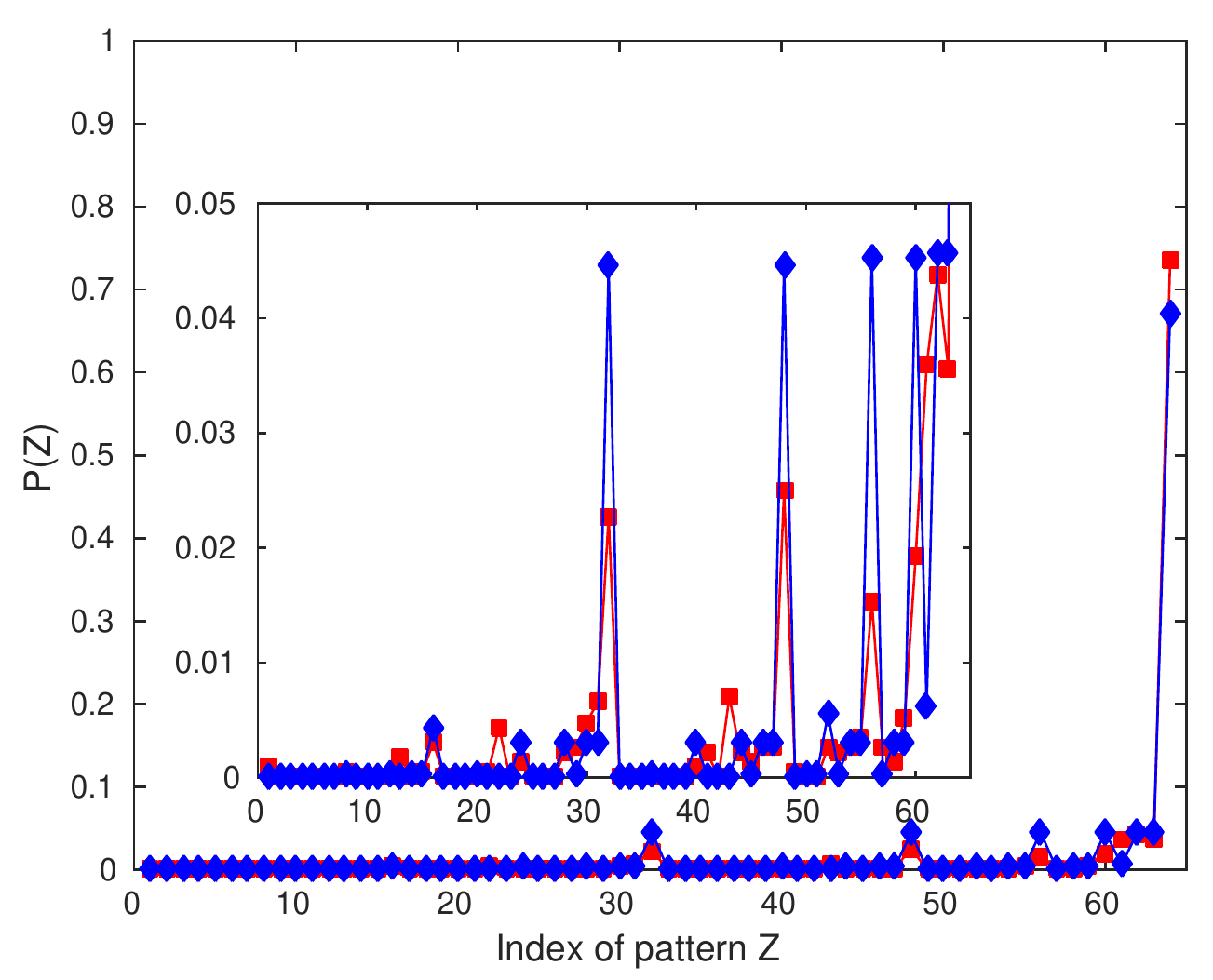}}\hspace{0.5cm}
\subfigure[(2,4)]{\includegraphics[scale=0.4]{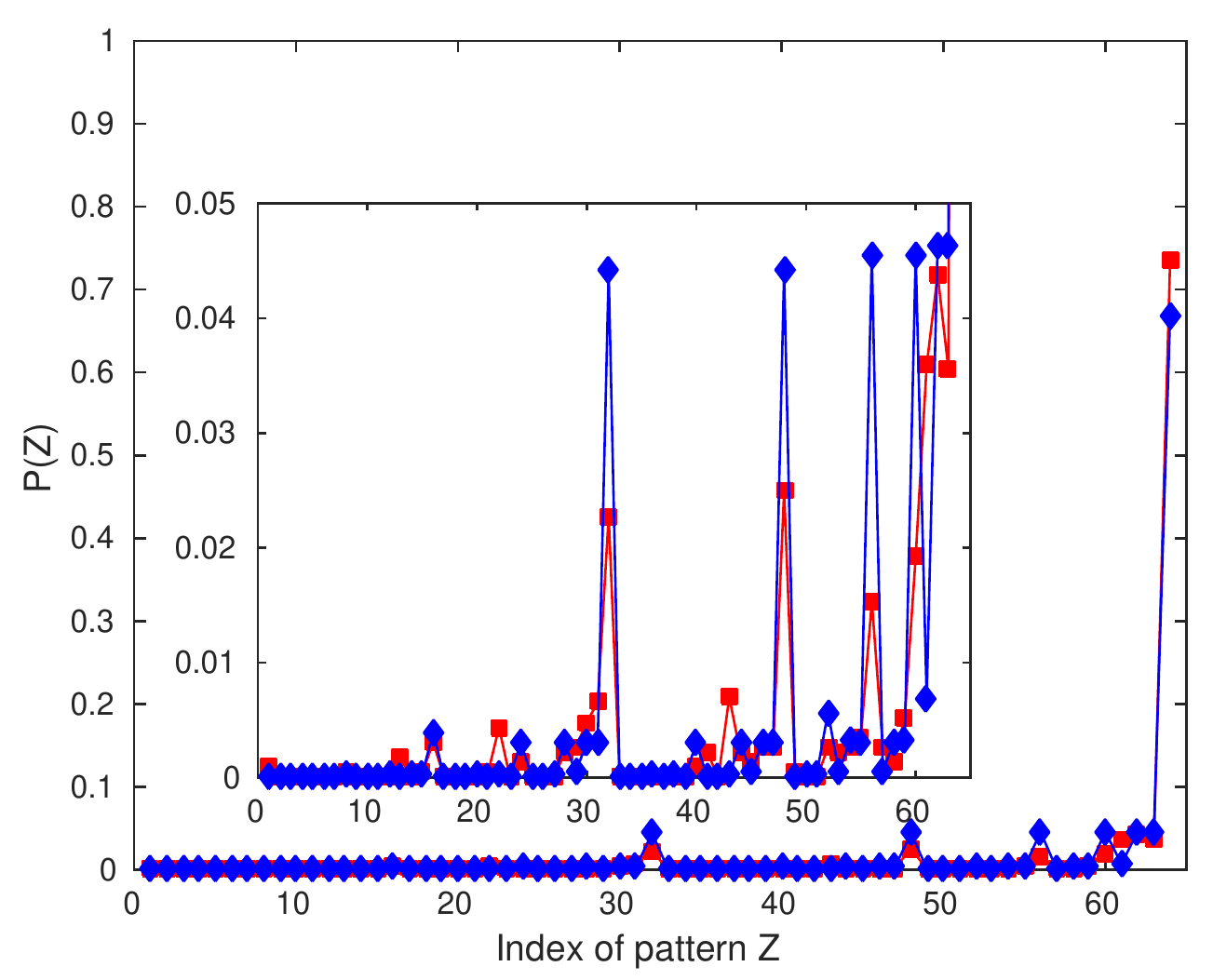}}
\caption{The figures show the predicted and the measured pattern distribution for all 16 models for mSat. The inset shows a zoomed in version of the distribution. \label{Fig:App:Prediction_mod1}}
\end{figure}
\pagebreak
\addtocounter{figure}{-1}
\begin{figure}[H]
\centering
\addtocounter{subfigure}{8}
\subfigure[(3,1)]{\includegraphics[scale=0.4]{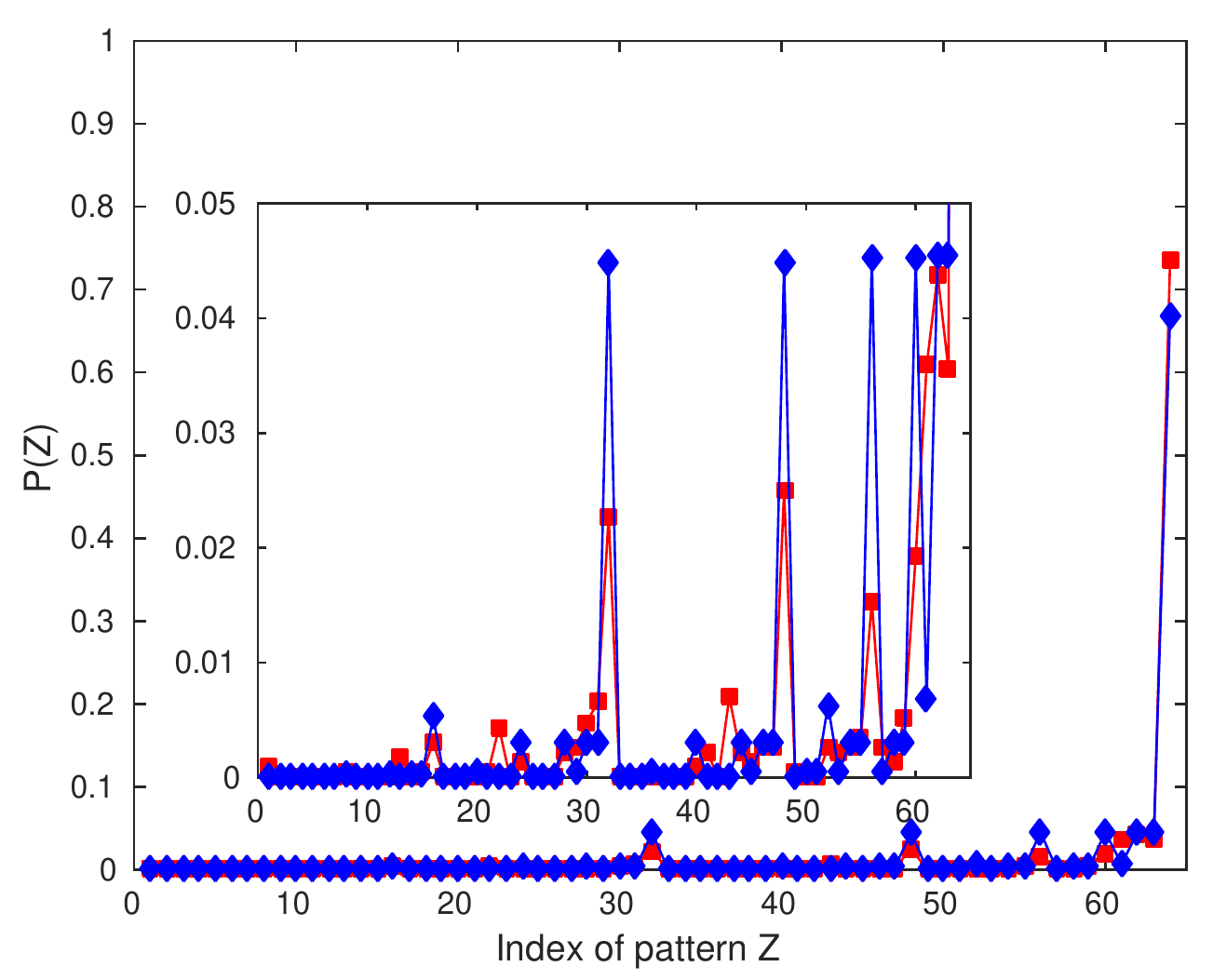}}\hspace{0.5cm}
\subfigure[(3,2)]{\includegraphics[scale=0.4]{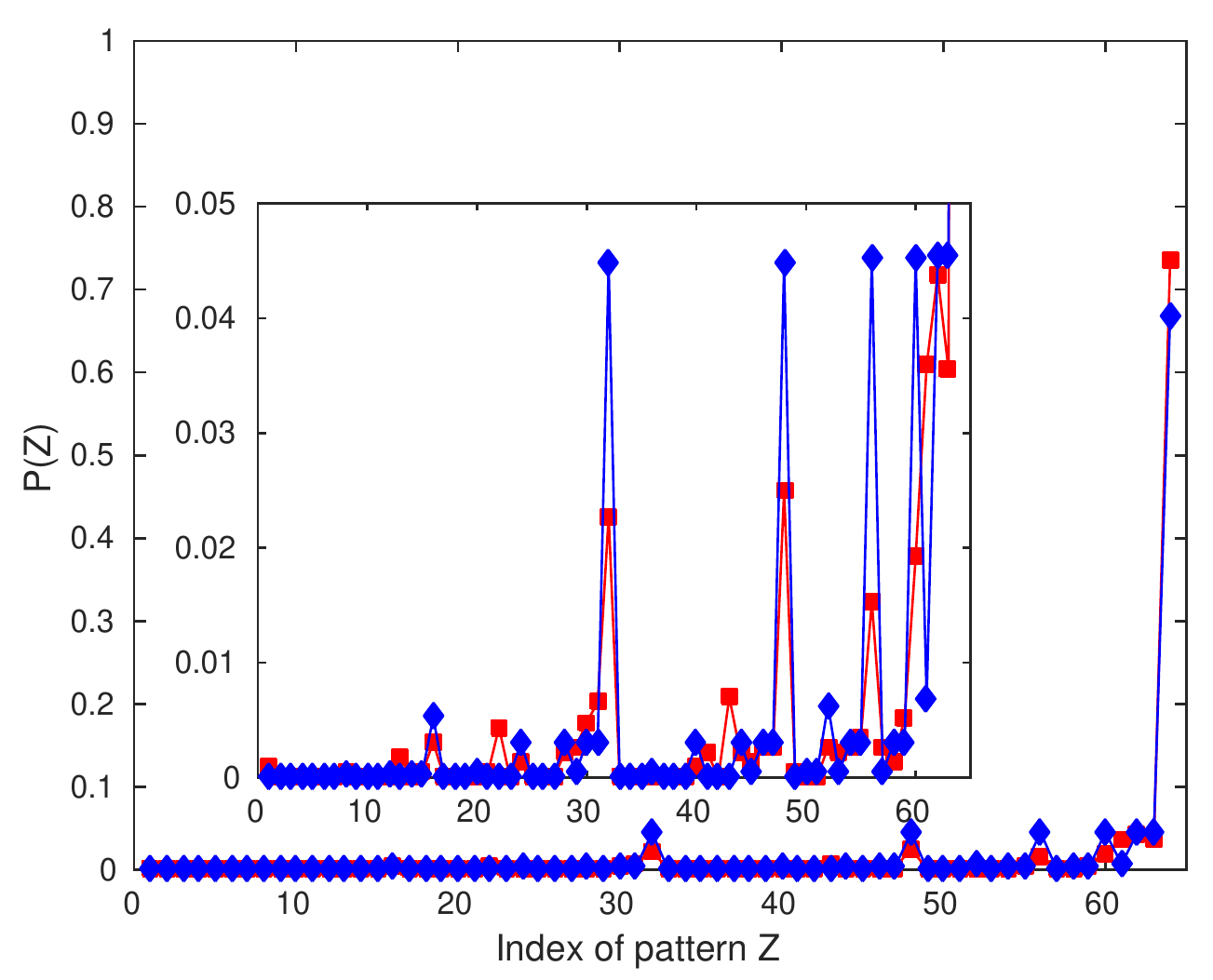}}\\[-0.2cm]
\subfigure[(3,3)]{\includegraphics[scale=0.4]{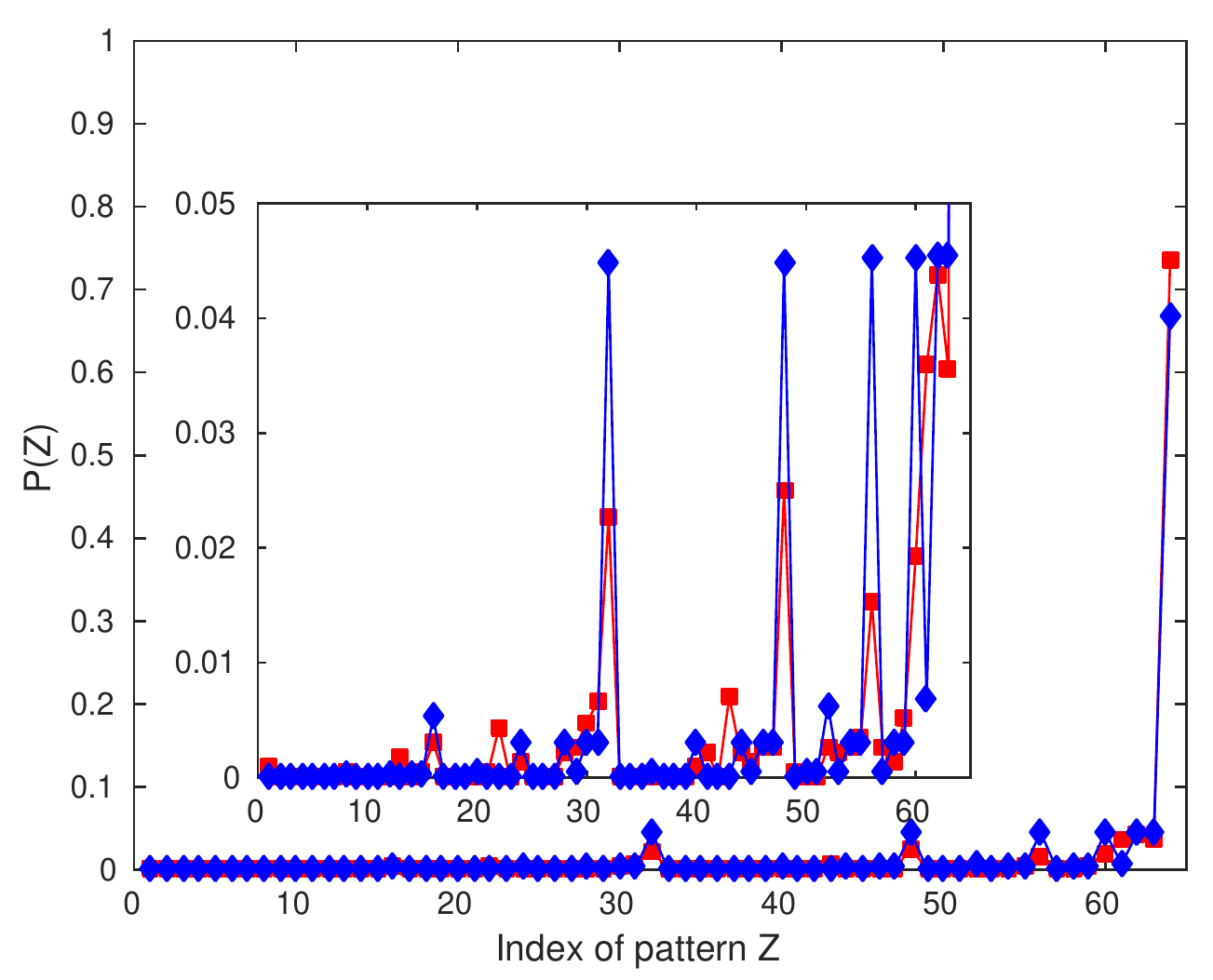}}\hspace{0.5cm}
\subfigure[(3,4)]{\includegraphics[scale=0.4]{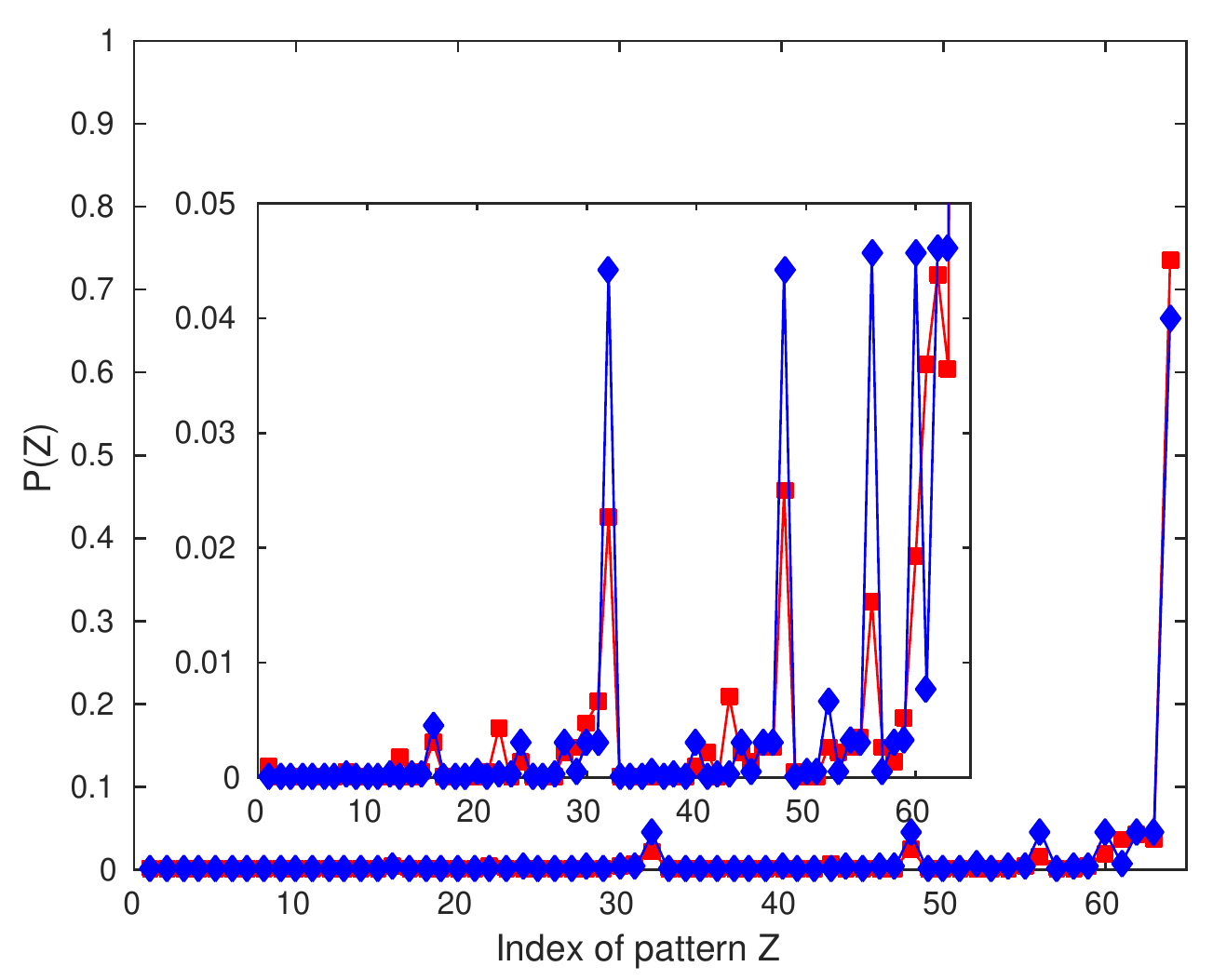}}\\[-0.2cm]
\subfigure[(4,1)]{\includegraphics[scale=0.4]{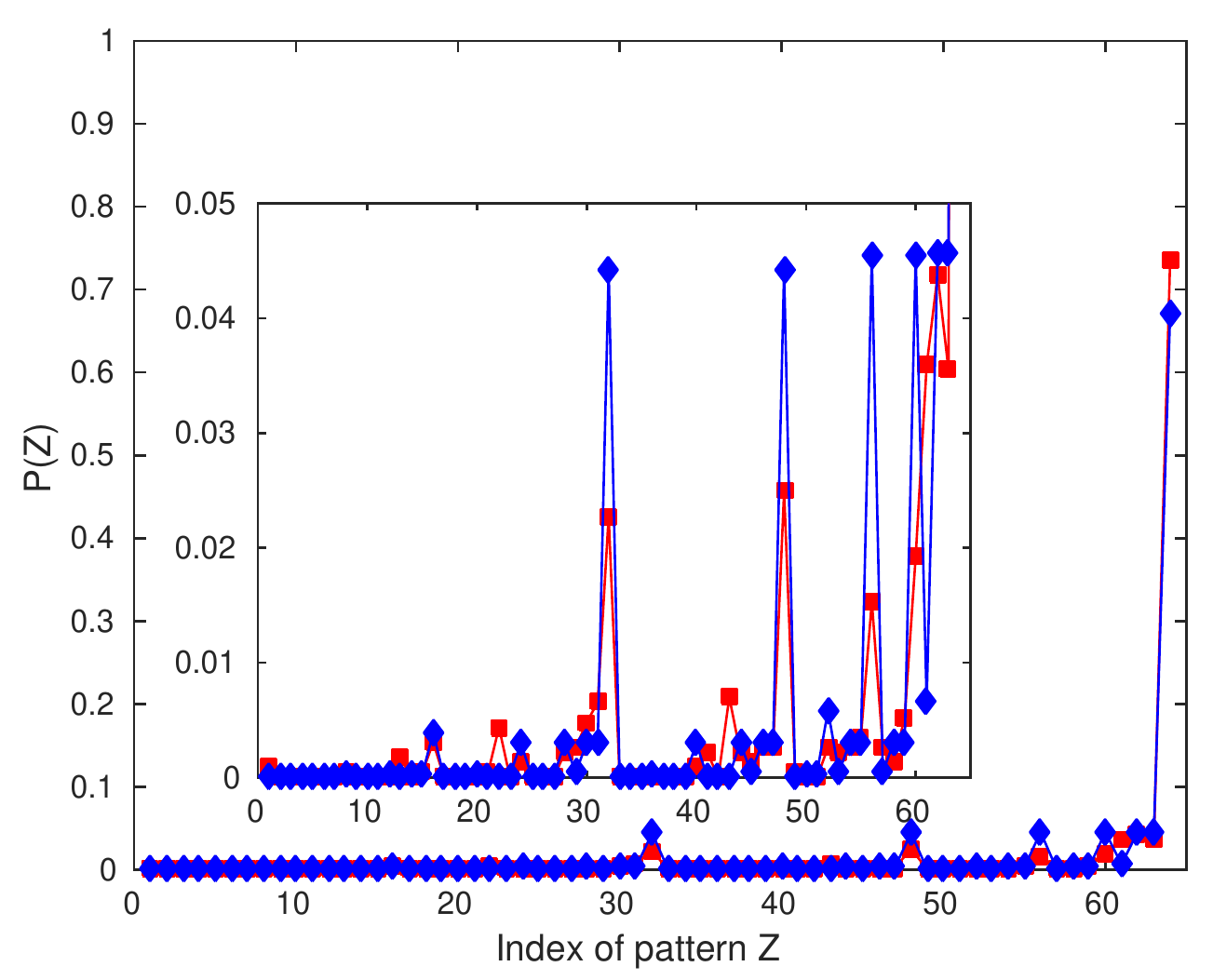}}\hspace{0.5cm}
\subfigure[(4,2)]{\includegraphics[scale=0.4]{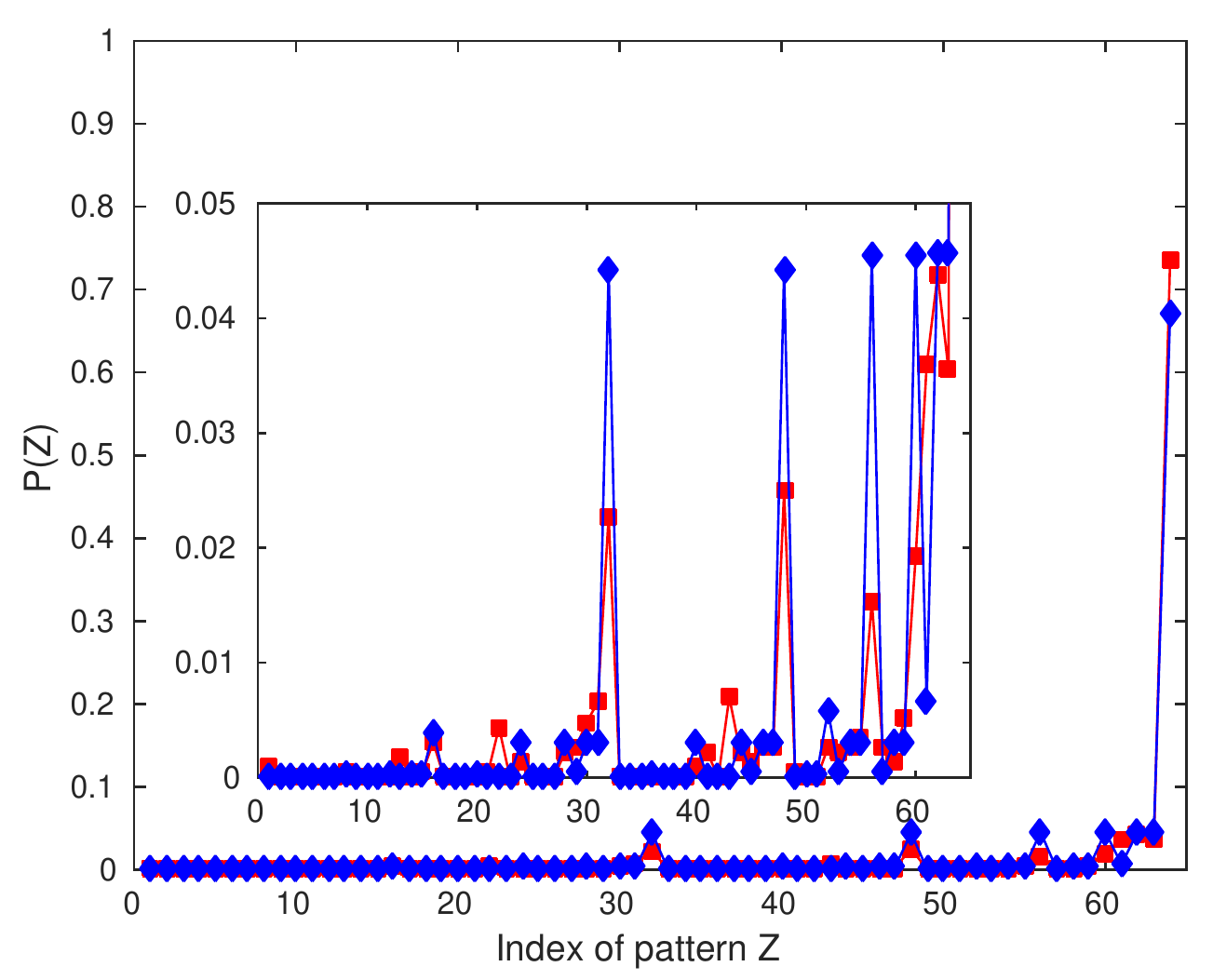}}\\[-0.2cm]
\subfigure[(4,3)]{\includegraphics[scale=0.4]{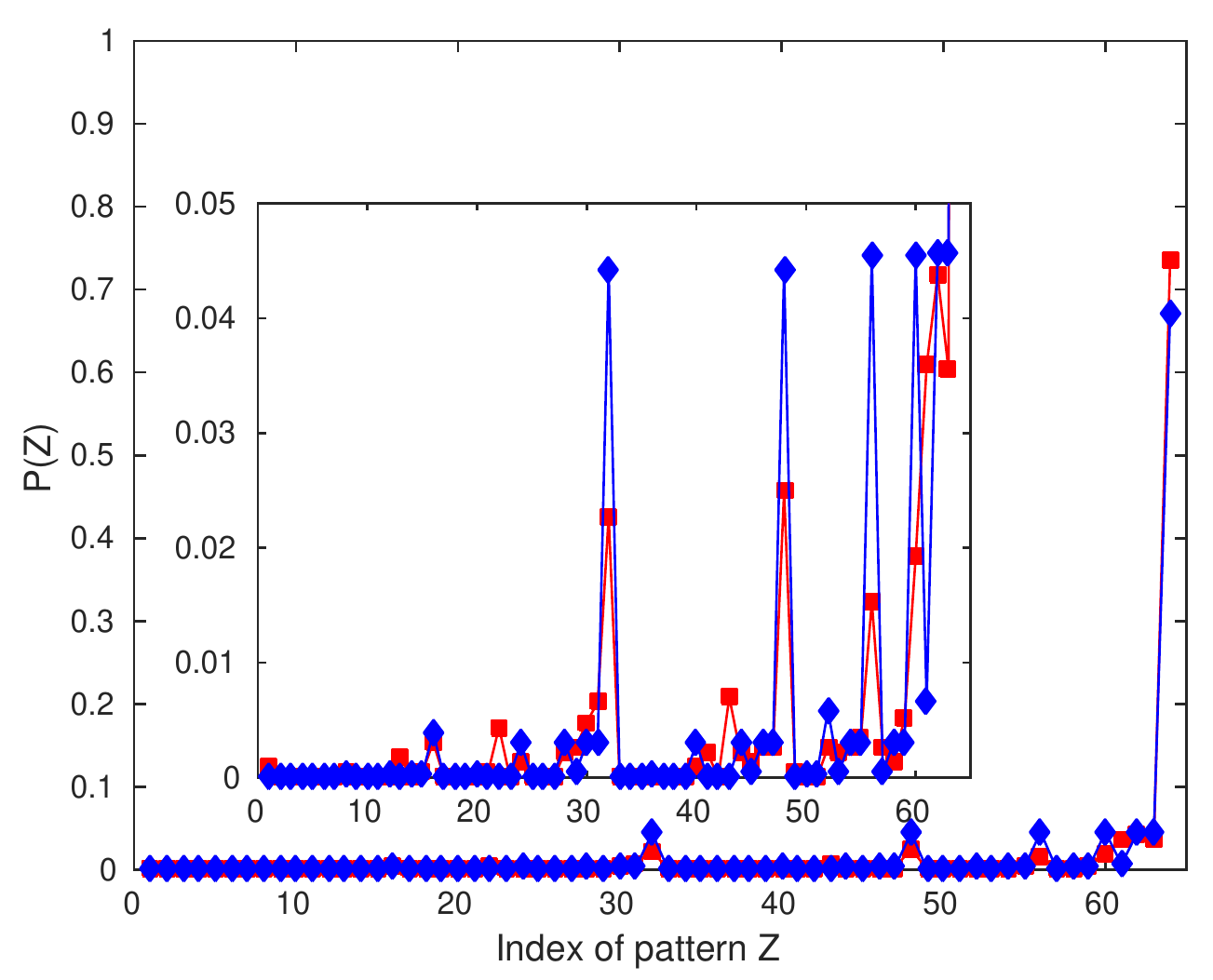}}\hspace{0.5cm}
\subfigure[(4,4)]{\includegraphics[scale=0.4]{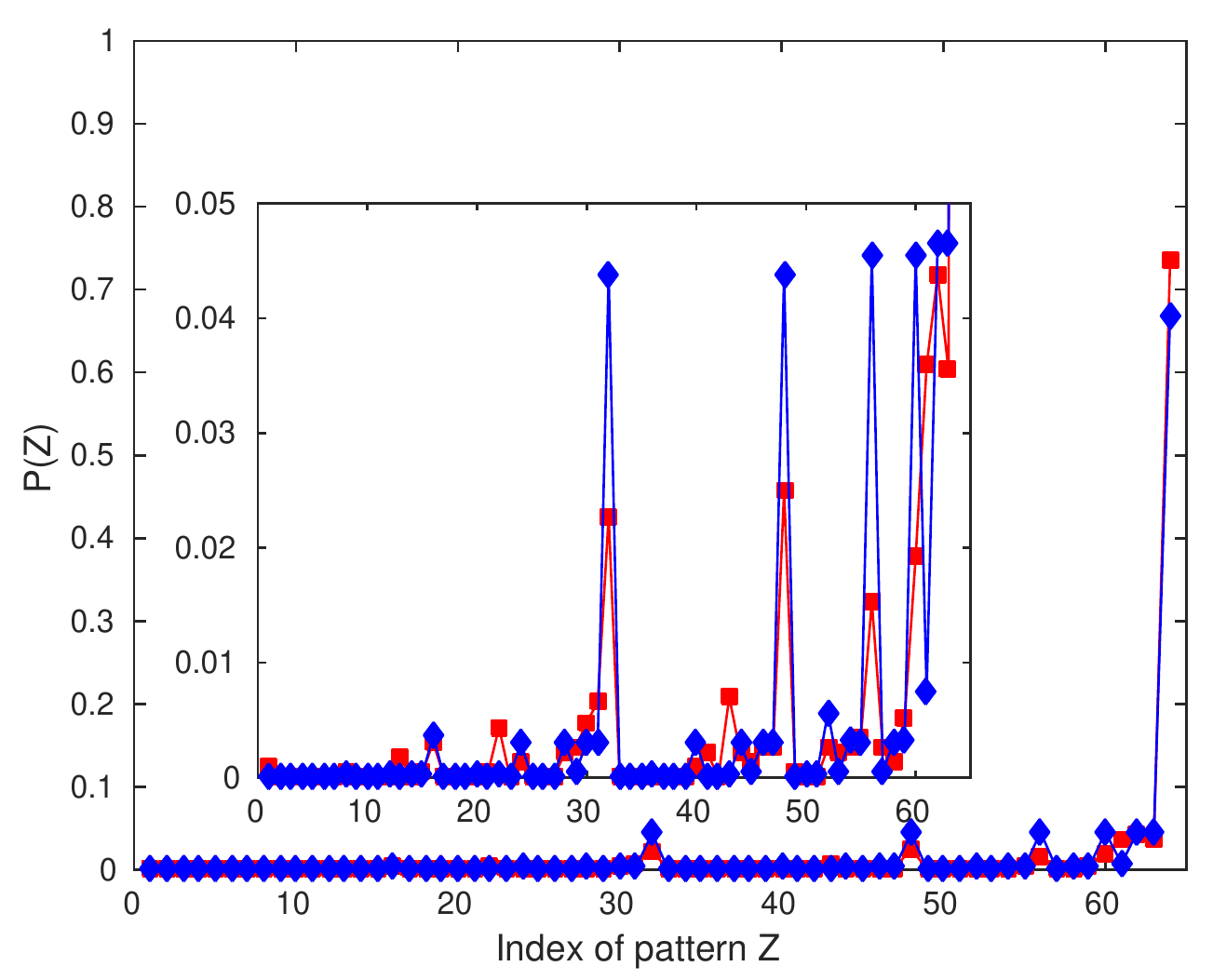}}
\caption{\textbf{(cont.)} The figures show the predicted and the measured pattern distribution for all 16 models for mSat. The inset shows a zoomed in version of the distribution.}
\end{figure}

\renewcommand{\arraystretch}{1.25}
\begin{table}[H]
\centering
\caption{Kullback-Leibler divergence $KL$ for the 16 models.\label{Tab:App:KL}}
\begin{tabular}{|c||c|c|c|c|}
\hline 
~~Model~~ & $(1,1)$ & $(1,2)$ & $(1,3)$ & $(1,4)$ \\
\hline
$KL$ & ~~$0.1398\pm0.0134$~~ & ~~$0.1398\pm0.0134$~~ & ~~$0.1398\pm0.0134$~~ & ~~$0.1337\pm0.0127$~~\\
\hline 
Model & $(2,1)$ & $(2,2)$ & $(2,3)$ & $(2,4)$ \\ 
\hline 
$KL$ & $0.1438\pm0.0137$ & $0.1439\pm0.0136$ & $0.1439\pm0.0137$ & $0.1374\pm0.0133$ \\ 
\hline 
Model & $(3,1)$ & $(3,2)$ & $(3,3)$ & $(3,4)$ \\ 
\hline 
$KL$ & $0.1399\pm0.0134$ & $0.1399\pm0.0134$ & $0.1398\pm0.0133$ & $0.1337\pm0.0127$ \\ 
\hline 
Model & $(4,1)$ & $(4,2)$ & $(4,3)$ & $(4,4)$ \\ 
\hline 
$KL$ & $0.1410\pm0.0137$ & $0.1411\pm0.0136$ & $0.1409\pm0.0135$ & $0.1349\pm0.0130$ \\ 
\hline 
\end{tabular}
\end{table}
\renewcommand{\arraystretch}{1}

\section{Related Work}
 In \cite{bonello2013bayesian} location- and neighbor-dependent models are proposed for single-stranded DNA
  methylation data in blood and tumor cells. 
 The (de-)methylation rates depend on the position of the CpG relative to the 3' or 5' end and/or on the methylation state of the left neighbor only. 
 The dependency is realized by the introduction of an additional parameter. 
 In our proposed models we use double-stranded DNA and can therefore include hemi-methylated sites and even distinguish on which strand the site is methylated. Furthermore we allow dependencies on both neighbors by introducing two different dependency parameters. 
 In contrast \cite{fu2010statistical} copes with the neighborhood dependency indirectly by allowing different parameter values for different sites. 
 In order to reduce the dimensionality of the parameter vector, a hierarchical model based on beta distributions is proposed. Another difference to our model is the distinction between de novo rates for parent and daughter strand. However, this can easily be included in future work. 
A density-dependent Markov model was proposed \cite{lacey2009modeling}. 
In this model, the probabilities of (de-)methylation events may depend on the methylation density in the CpG neighborhood. 
In addition,  a neighboring sites model has been developed, in which the probabilities for a given site are directly influenced by the states of neighboring sites to the left and right \cite{lacey2009modeling}. 
When these models were tested on double-stranded methylation patterns from two distinct tandem repeat regions in a collection of ovarian carcinomas, the density-dependent and neighboring sites models were superior to   independent models in generating statistically similar samples.  
 Although this model also includes the dependence on the methylation state on the left and right neighbor for double-stranded DNA the approach is different. The transition probabilities of 
 the neighbor-independent model are transformed into a   transition probability of a neighbor-dependent model
 by introducing only one additional parameter. The state of the left and right neighbor are taken into account
  by exponentiating this parameter by some norm. 
 In addition, this approach does not allow the intuitive interpretation of the dependency parameter.

\section{Conclusion}
 We proposed a set of stochastic models for the formation and modification of methylation patterns 
over time. These models take into account the state of the CpG sites in the spatial neighborhood and 
 allow to describe different hypotheses about the underlying mechanisms  of
 methyltransferases adding methyl groups at CpG sites. We used knockout data from bisulfite sequencing  
 at 
 several loci to learn the efficiencies at which these enzymes perform methylation.
 By combining these efficiencies, we accurately predicted the probability distribution of the patterns in the
 wild-type. 
Moreover, we found that in all cases the models predict values for the dependency parameters
$\psi_L$ and $\psi_R$ close to 1 and therefore independence of methylation for the Dnmt3a/b DKO 
meaning that Dnmt1 methylates CpGs independent of the methylation of neighboring CpGs. 
For Dnmt3a/b on the other hand we could identify dependencies on the neighboring CpGs.
Both findings are in accordance with   current existing mechanistic models:
 Dnmt1 reliably copies the methylation from the template strand to maintain the distinct methylation patterns, whereas Dnmt3a/b try to establish and keep a certain amount of CpG methylation at a given loci.
Interestingly, our 
models only suggest dependencies of de novo methylation activity on the CpGs in the 5' neighborhood. 
 This indicates that Dnmt3a and Dnmt3b show a preference to methylate CpGs in a 5' to 3' direction and could point towards a processive or cooperative behavior of these enzymes like recently described  in \emph{in vitro} experiments
\cite{emperle2014cooperative,holz2010inherent}.
Compared to a neighborhood independent model with $\psi_L=\psi_R=1$, a neighborhood dependent model shows
better predictions and furthermore allows to investigate (possible) connections of adjacent CpGs and their methylation states.


%

As future work, we plan to investigate models in which we distinguish between the actions of Dnmt3a and Dnmt3b
and in which we allow a diagonal dependency for de novo methylation, i.e., a dependency on the state of neighboring CpGs on the opposite strand.  Moreover, we will design models that take into account the 
number of base pairs between adjacent CpG sites. 
To investigate a potential impact of oxidized cytosine forms on the methylation at neighboring CpG sites we further plan to include the CpG states 5hmC, 5fC and 5caC   in our model.


\begin{thebibliography}{10}
\providecommand{\url}[1]{\texttt{#1}}
\providecommand{\urlprefix}{URL }

\bibitem{aijo2016probabilistic}
{\"A}ij{\"o}, T., Huang, Y., Mannerstr{\"o}m, H., Chavez, L., Tsagaratou, A.,
  Rao, A., L{\"a}hdesm{\"a}ki, H.: A probabilistic generative model for
  quantification of {DNA} modifications enables analysis of demethylation
  pathways. Genome Biology  17(1), ~49 (2016)

\bibitem{arand2012vivo}
Arand, J., Spieler, D., Karius, T., Branco, M.R., Meilinger, D., Meissner, A.,
  Jenuwein, T., Xu, G., Leonhardt, H., Wolf, V., et~al.: In vivo control of
  {CpG} and non-{CpG} {DNA} methylation by {DNA} methyltransferases. PLoS Genet
   8(6),  e1002750 (2012)

\bibitem{baubec2015genomic}
Baubec, T., Colombo, D.F., Wirbelauer, C., Schmidt, J., Burger, L., Krebs,
  A.R., Akalin, A., Sch{\"u}beler, D.: Genomic profiling of {DNA}
  methyltransferases reveals a role for {DNMT3B} in genic methylation. Nature
  520(7546),  243--247 (2015)

\bibitem{bonello2013bayesian}
Bonello, N., Sampson, J., Burn, J., Wilson, I.J., McGrown, G., Margison, G.P.,
  Thorncroft, M., Crossbie, P., Povey, A.C., Santibanez-Koref, M., et~al.:
  Bayesian inference supports a location and neighbour-dependent model of {DNA}
  methylation propagation at the {MGMT} gene promoter in lung tumours. Journal
  of Theoretical Biology  336,  87--95 (2013)

\bibitem{emperle2014cooperative}
Emperle, M., Rajavelu, A., Reinhardt, R., Jurkowska, R.Z., Jeltsch, A.:
  Cooperative {DNA} binding and protein/{DNA} fiber formation increases the
  activity of the {Dnmt3a} {DNA} methyltransferase. Journal of Biological
  Chemistry  289(43),  29602--29613 (2014)

\bibitem{fu2010statistical}
Fu, A.Q., Genereux, D.P., St{\"o}ger, R., Laird, C.D., Stephens, M.:
  Statistical inference of transmission fidelity of {DNA} methylation patterns
  over somatic cell divisions in mammals. The Annals of Applied Statistics
  4(2),  871 (2010)

\bibitem{genereux2005population}
Genereux, D.P., Miner, B.E., Bergstrom, C.T., Laird, C.D.: A
  population-epigenetic model to infer site-specific methylation rates from
  double-stranded {DNA} methylation patterns. PNAS  102(16),  5802--5807 (2005)

\bibitem{giehr2016}
Giehr, P., Kyriakopoulos, C., Ficz, G., Wolf, V., Walter, J.: The {Influence}
  of {Hydroxylation} on {Maintaining} {CpG} {Methylation} {Patterns}: A
  {Hidden} {Markov} {Model} {Approach}. PLoS Comput Biol  12(5),  e1004905
  (2016)

\bibitem{gowher2002molecular}
Gowher, H., Jeltsch, A.: Molecular enzymology of the catalytic domains of the
  {D}nmt3a and {D}nmt3b {DNA} methyltransferases. Journal of Biological
  Chemistry  277(23),  20409--20414 (2002)

\bibitem{hermann2004dnmt1}
Hermann, A., Goyal, R., Jeltsch, A.: The {Dnmt1}
  {DNA}-(cytosine-c5)-methyl\-transferase methylates {DNA} processively with
  high preference for hemimethylated target sites. Journal of Biological
  Chemistry  279(46),  48350--48359 (2004)

\bibitem{holz2010inherent}
Holz-Schietinger, C., Reich, N.O.: The inherent processivity of the human de
  novo methyltransferase 3{A} ({DNMT3A}) is enhanced by {DNMT3L}. Journal of
  Biological Chemistry  285(38),  29091--29100 (2010)

\bibitem{kapourani2016higher}
Kapourani, C.A., Sanguinetti, G.: Higher order methylation features for
  clustering and prediction in epigenomic studies. Bioinformatics  32(17),
  i405--i412 (2016)

\bibitem{kyriakopoulos2016h}
Kyriakopoulos, C., Giehr, P., Wolf, V.: {H(O)TA}: estimation of {DNA}
  methylation and hydroxylation levels and efficiencies from time course data.
  Bioinformatics  (2017), to appear.

\bibitem{lacey2009modeling}
Lacey, M.R., Ehrlich, M., et~al.: Modeling dependence in methylation patterns
  with application to ovarian carcinomas. Stat Appl Genet Mol Biol  8(1), ~40
  (2009)

\bibitem{laird2004hairpin}
Laird, C.D., Pleasant, N.D., Clark, A.D., Sneeden, J.L., Hassan, K.A., Manley,
  N.C., Vary, J.C., Morgan, T., Hansen, R.S., St{\"o}ger, R.: Hairpin-bisulfite
  {PCR}: assessing epigenetic methylation patterns on complementary strands of
  individual {DNA} molecules. PNAS  101(1),  204--209 (2004)

\bibitem{norvil2016dnmt3b}
Norvil, A.B., Petell, C.J., Alabdi, L., Wu, L., Rossie, S., Gowher, H.:
  {D}nmt3b {M}ethylates {DNA} by a {N}oncooperative {M}echanism, and {I}ts
  {A}ctivity {I}s {U}naffected by {M}anipulations at the {P}redicted {D}imer
  {I}nterface. Biochemistry  (2016)

\bibitem{okano1999dna}
Okano, M., Bell, D.W., Haber, D.A., Li, E.: {DNA} methyltransferases {D}nmt3a
  and {D}nmt3b are essential for de novo methylation and mammalian development.
  Cell  99(3),  247--257 (1999)

\bibitem{otto1990dna}
Otto, S.P., Walbot, V.: {DNA} methylation in eukaryotes: kinetics of
  demethylation and de novo methylation during the life cycle. Genetics
  124(2),  429--437 (1990)

\bibitem{sontag2006dynamics}
Sontag, L.B., Lorincz, M.C., Luebeck, E.G.: Dynamics, stability and inheritance
  of somatic {DNA} methylation imprints. Journal of Theoretical Biology
  242(4),  890--899 (2006)

\bibitem{suzuki2008dna}
Suzuki, M.M., Bird, A.: {DNA} methylation landscapes: provocative insights from
  epigenomics. Nature Reviews Genetics  9(6),  465--476 (2008)

\end{thebibliography}
\end{document}